\documentclass[fleqn,usenatbib]{mnras}

\usepackage[T1]{fontenc}
\usepackage{ae,aecompl}

\usepackage{graphicx}	
\usepackage{amsmath}	
\usepackage{amssymb}	
\usepackage{mathtools}
\usepackage{hyperref}
\usepackage{xspace}	
\usepackage{xcolor}	
\usepackage{euscript}
\usepackage{pdflscape}
\usepackage{afterpage}
\usepackage{placeins}
\usepackage{bm}
\usepackage{ dsfont }
\usepackage{listings}
\usepackage{soul}
\usepackage{bbm}

\usepackage{newtxtext,newtxmath}

\definecolor{codegreen}{rgb}{0,0.6,0}
\definecolor{codegray}{rgb}{0.5,0.5,0.5}
\definecolor{codepurple}{rgb}{0.58,0,0.82}
\definecolor{backcolour}{rgb}{0.95,0.95,0.92}

\lstdefinestyle{python}{
    backgroundcolor=\color{backcolour},   
    commentstyle=\color{codegreen},
    keywordstyle=\color{magenta},
    numberstyle=\tiny\color{codegray},
    stringstyle=\color{codepurple},
    basicstyle=\ttfamily\footnotesize,
    breakatwhitespace=false,         
    breaklines=true,                 
    captionpos=b,                    
    keepspaces=true,                 
    numbers=left,                    
    numbersep=5pt,                  
    showspaces=false,                
    showstringspaces=false,
    showtabs=false,                  
    tabsize=2,
    upquote=true
}
\definecolor{dkgreen}{rgb}{0,0.6,0}
\definecolor{gray}{rgb}{0.5,0.5,0.5}
\definecolor{mauve}{rgb}{0.58,0,0.82}
\newcommand{\gaia}{{\it Gaia}\xspace}
\newcommand{\prob}{\mathrm{P}}

\defcitealias{PaperI}{Paper~I}
\defcitealias{PaperII}{Paper~II}

\title[\textit{Gaia} Astrometry and RVS Selection Functions]{Completeness of the \textit{Gaia}-verse V: Astrometry and Radial Velocity sample selection functions in \textit{Gaia} EDR3}

\author[A. Everall and D. Boubert]{ 
	Andrew Everall$^{1}$\thanks{E-mail: asfe2@cam.ac.uk}
	and
	Douglas Boubert$^{2,3}$
	\\
$^{1}$Institute of Astronomy, University of Cambridge, Madingley Road, Cambridge CB3 0HA, UK\\
	$^{2}$Magdalen College, University of Oxford, High Street, Oxford OX1 4AU, UK\\
	$^{3}$Rudolf Peierls Centre for Theoretical Physics, Clarendon Laboratory, Parks Road, Oxford OX1 3PU, UK\\
}

\date{Accepted XXX. Received YYY; in original form ZZZ}

\pubyear{}

\begin{document}
\label{firstpage}
\pagerange{\pageref{firstpage}--\pageref{lastpage}}
\maketitle

\begin{abstract}
We risk reaching false scientific conclusions if we test our physical theories against subsets of the \textit{Gaia} catalogue without correcting for the biased process by which stars make it into our sample. In this paper we produce selection functions for three \textit{Gaia} science samples to enable the community to correct for this bias.
We estimate the probability that a source in \textit{Gaia} EDR3 will have i) a reported parallax and proper motion, ii) an astrometric renormalised unit weight error below 1.4, or iii) a reported radial velocity. These selection functions are estimated as a function of $G$-band apparent magnitude and position on the sky, with the latter two also being dependent on $G-G_\mathrm{RP}$ colour. The inferred selection functions have a non-trivial dependence on these observable parameters, demonstrating the importance of empirically estimating selection functions. We also produce a simple estimate for the selection function of the full \textit{Gaia} EDR3 source catalogue to be used with the subset selection functions. We make all of our selection functions easily accessible through the GitHub repository \textsc{selectionfunctions}.
\end{abstract}

\begin{keywords}
	stars: statistics, Galaxy: kinematics and dynamics, Galaxy: stellar content, methods: data analysis, methods: statistical
\end{keywords}



\section{Introduction}

As observational astronomers, we use data from observatories such as the \gaia satellite \citep{Prusti2016} to test our physical theories about stars, galaxies and the universe. However, to reliably test these theories we must understand our data and account for regions of parameter space which couldn't be observed. No observatory paints a complete picture of the universe and in order to fill in the gaps with our models, we need to know what could have been observed and what could not. We achieve this through a selection function. For any catalogue of observed sources the selection function tells us "Given a real or hypothetical object, what is the probability that the object would have been successfully observed and included in the catalogue?". A detailed exposition of selection functions and their applications is provided in \citep{Rix2021}.

The \gaia DR2 selection function, modelled in \citet{CoGII}, has enabled us to answer the question "What is the probability that a source was observed and recorded in the \gaia DR2 source catalogue?". However, additional information is provided for subsets of objects in the \gaia source catalogue which are vital for answering certain scientific questions. For example, the sample with measured parallax and proper motion \citep[][]{Lindegren2018, Lindegren2021} is enabling us to study and understand the structure and kinematics of the Milky Way, whilst the sample with radial velocities \citep[][]{Katz2018} includes the final dimension of kinematic data for a smaller, but no less impressive, set of objects.


Usually selection functions are estimated by comparing the sample to a more complete source catalogue. For the \gaia source catalogue, there is no more complete sample to compare against. Therefore we had to use our understanding of the observing strategy of \gaia to model the selection function \citep{CoGII}. However, for subsets of \gaia, the source catalogue provides a sample which we know is more complete in all areas of parameter space. Previous attempts at selection functions for samples where a more complete catalogue is available involve taking the ratio of objects in the subset to source catalogue in carefully chosen on-sky position-colour-apparent magnitude bins \citep[e.g. ]{Bovy2012,Wojno2017,Mints2019}. This requires a selection function which can be defined discretely with enough objects in each bin of the subset and source catalogues such that the ratio is approximately equal to the expected completeness. This is not the case for subsets of \gaia which have complex selection criteria within the processing pipeline relating to satellite performance and the scanning law \citep{CoGI,CoGIII}. 

A strong example of this is the \gaia DR2 radial velocity catalogue (which we will address in this paper). \citet{Rybizki2021RVS} attempted to evaluate the selection function for this sample relative to the \gaia source catalogue using number count ratios in colour-magnitude-sky bins. However, as they demonstrate, the selection function has a strong dependence on sky position over small scales due to the scanning law, crowding limitations and the Initial Gaia Source List \citep[IGSL, ][]{Smart2014}. Poisson noise prevented them from going to sufficient resolution to model this structure without running out of objects in \gaia.

\citet{Everall2020} went beyond simple count ratios by modelling the source catalogue and subset CMDs as continuous Poisson count processes with Gaussian Mixture Models. Whilst this model was well-suited to multi-fibre spectrographs with well defined fields, a significant improvement was needed to fit the all-sky selection function of \gaia catalogues.

This major step forward has come from \citet{Boubert2021} which presents a general approach to estimating selection functions for subsets of  catalogues. They model the selection function as a sum of well-localised spherical needlets on the sky with weights drawn from a Gaussian Process across colour and apparent magnitude. They also use the formally correct Binomial likelihood function to fit the model.

In this work we will apply the \citet{Boubert2021} model to three subsets of \gaia EDR3: 
\begin{enumerate}
    \item Astrometry: the sample with published parallax and proper motion.\footnote{In \gaia EDR3 this is the combined sample with 5D or 6D astrometric solutions. We note that sources without parallax and proper motions in \gaia still received 2D astrometric solutions however we refer to the sample with 5/6D solutions as the astrometry sample for brevity.}
    \item $\mathrm{RUWE}$: Renormalised unit weight error specifies the goodness of fit of an astrometric solution and is often used to filter out poorly fit sources. We provide selection functions for the sample with $\mathrm{RUWE}<1.4$
    \item RVS: The sample with published radial velocities from the radial velocity spectrograph (RVS).
\end{enumerate}

We'll provide a brief review of the model developed by \citet{Boubert2021} in Section~\ref{sec:method} and explain how binomial statistics are used to fit the likelihood function and test the results in Section~\ref{sec:binomial}. Readers primarily interested in the results may wish to skip this and go directly to Section~\ref{sec:data} where the selected samples are described. We present our results in Section~\ref{sec:results}. Before concluding we also discuss other \gaia catalogues which have not been investigated here due to additional complexities but which should be a key goal for future work in this field.

\section{Methodology}
\label{sec:method}

An in-depth description of the method for constructing sub-sample selection functions for astronomical surveys is given in \citet{Boubert2021}. Here we provide a brief summary of the key points which are used in this paper.

The aim of this work is to estimate the selection probabilities of subsets of the \gaia EDR3 relative to the source catalogue. e.g. `What is the probability that a source in \gaia EDR3 has parallax and proper motion?'. This selection function may be written down as
\begin{equation}
    \prob(\mathcal{S}_\mathrm{subset} \,|\, \mathcal{S}_\mathrm{source}, \mathbf{y})
    \label{eq:sfrel}
\end{equation}
where $\mathbf{y}$ are the observables (or functions of observables) over which the selection function is defined and `source' refers to the more complete catalogue, in this case the \gaia source catalogue. The probability of any object being included in the subset is 
\begin{equation}
    \prob(\mathcal{S}_\mathrm{subset} \,|\, \mathbf{y}) = \prob(\mathcal{S}_\mathrm{subset} \,|\, \mathcal{S}_\mathrm{source}, \mathbf{y})\,\cdot\,\prob(\mathcal{S}_\mathrm{source} \,|\, \mathbf{y})
    \label{eq:schild}
\end{equation}
where the source catalogue selection function, $\prob(\mathcal{S}_\mathrm{source} \,|\, \mathbf{y})$, is the probability that a source is included in the \gaia source catalogue. The source catalogue selection function for \gaia DR2 has previously been estimated by \citet{CoGII} and a significantly improved version for the EDR3 catalogue is being developed \citep{CoGVI,CoGVII}. For users who wish to apply these models in their own work sooner, we provide an initial estimate of the \gaia EDR3 source catalogue selection function in Appendix~\ref{app:edr3}, however, this will be superseded by \citet{CoGVI,CoGVII} when published.

All selection functions estimated here will be over the variables $\mathbf{y}=(l,b,G)$ with the RUWE and RVS selection functions additionally a function of $G-G_\mathrm{RP}$. These are the dominant variables in selecting sources to enter the samples. The scanning law produces complex selection patterns across the sky generating a heavy dependence on $(l,b)$. Whether individual observations are used in the \gaia data is dependent on the on-board estimated apparent magnitude of the source (scientific measurements of a source are only made if $G_\mathrm{onboard}<20.7$, \citet{Prusti2016}), for which $G$ is a reasonable proxy. The publication of measured radial velocity for any source in \gaia is contingent on an estimated RVS magnitude ($G_\mathrm{RVS}$) calculated using the IGSL \citep[See Section 2.1 in ][]{Sartoretti2018}. As will be discussed later, $G_\mathrm{RVS}$ is more similar to $G_\mathrm{RP}$ than $G$ which means the magnitude limit in $G$ will be a function of $G-G_\mathrm{RP}$. Finally, the selection criteria for the RVS sample is explicitly dependent on source temperature and RUWE is implicitly dependent on colour. Making our selection function dependent on $G-G_\mathrm{RP}$ allows us to capture these dependencies. We choose to use $G-G_\mathrm{RP}$ rather than $G_\mathrm{BP}-G_\mathrm{RP}$ due to calibration issues at the faint end of $G_\mathrm{BP}$ \citep{Riello2021}. The drawback of this is that $G_\mathrm{RP}$ uses larger spatial windows than $G$ which means that our colours will be extremely red for extended sources with high excess flux \citep[see Section~9.4][]{Riello2021}.

Whilst we only consider magnitude, colour and position on the sky in this work, we are also conscious that \gaia is heavily limited in crowded regions of the sky such as the Milky Way bulge, LMC and globular clusters. We will discuss why we have not directly used crowding as a selection function variable and the implications of this in Section~\ref{sec:discussion}.

The selection function, as described in \citet{Boubert2021}, is composed of a sum of spherical needlets across the sky with coefficients subject to a Gaussian Process prior in apparent magnitude and colour. This is fit to the data using a Binomial likelihood function in HEALPix-apparent magnitude-colour bins. In the following subsections, we'll briefly describe the maths of each of these components.

\subsection{Logit probability}

Our model will estimate the probability, $q \in [0,1]$ that an object will be included in the subset given the observables and that it's in the source catalogue. Our model is defined in an infinite domain so we directly model $x \in [-\infty,\infty]$, the logit-transformed probability
\begin{equation}
    x = \mathrm{logit}(q) \equiv \log\left(\frac{q}{1-q}\right),
\end{equation}
with $x$ sometimes being referred to as the log-odds ratio. This is then inverse transformed to retrieve the selection probability
\begin{equation}
    q = \mathrm{logit}^{-1}(x) = \frac{1}{1+\exp(-x)}
\end{equation}
which is also referred to as the `expit' function. All figures in the paper are given in terms of $x$ but we provide values of $q$ in the axes to help with interpretation.

\subsection{Needlets}

We use a smooth spatial model rather than estimating individual bin probabilities in order to avoid being dominated by noisy data.

Spherical harmonics provide an orthonormal basis for functions on the sphere and so are commonly-used to model distributions over a spherical surface. 
However, since individual spherical harmonics are not localised on the sky, a small change in the function at one location changes the coefficients of all the harmonics. This can be overcome by using a convolution of spherical harmonics called a `needlet' \citep{Marinucci2008}
\begin{equation}
    \psi_{j\kappa} = \sqrt{\lambda_{j\kappa}} \sum_{
    \ell=0}^{\ell_\mathrm{max}} b_{\ell}(j) \frac{2\ell+1}{4\pi} P_{\ell}(\cos(\phi_{j\kappa}(l,b))),
\end{equation}
where $P_{\ell}$ are the Legendre polynomials and $\phi_{j\kappa}(l,b)$ is the great arc separation between the coordinates $(l,b)$ and the Needlet centre. Localisation of the Needlets is achieved through the window function, $b(j)$. For this we use the `Chi-square' Needlets\footnote{Referred to in those works as `Mexican' Needlets.} described in \citet{Geller2009} and \citet{Scodeller2011},
\begin{equation}
    b_{\ell}(j|B,\nu) = \left(\frac{j}{B^j}\right)^{2\nu} \exp\left(-\frac{\ell^2}{B^{2j}}\right),
    \label{eq:window}
\end{equation}
where $B$ and $\nu$ are free parameters which canonically take the values two and one respectively. In order to satisfy the reconstruction formula for spherical Needlets, the window function must satisfy \citep{Baldi2006}
\begin{equation}
    \sum_{j=0}^{\infty} b(j)^2 \equiv 1.
\end{equation}
This is not generally true for the window function in Eq.~\ref{eq:window} so we numerically renormalise the window functions for each $\ell$ summing for $j$ up to $1000$ to guarantee the relation is satisfied.

The needlets are centered on HEALPix pixels \citep{Gorski2005} such that
\begin{equation}
    x(l,b) = \sum_{j=0}^{j_\mathrm{max}}\sum_{\kappa=0}^{N_j} \,\beta_{j\kappa}\,\psi_{j\kappa}(l,b)
    \label{eq:x_model}
\end{equation}
where $j_\mathrm{max}$ is the maximum HEALPix level used and $N_j = 12 \times 2^{2j}$ is the number of pixels in a given HEALPix level. The selection probabilities as a function of position on the sky are then given by $q = \mathrm{logit}^{-1}(x)$.

\citet{Boubert2021} provides a more detailed introduction to needlets and their Fig.~1 demonstrates how needlets appear on the sphere.

Needlets do not include an $\ell=0$ mode which corresponds a uniform contribution across the sky. We manually include this in our model as the $j=-1$ component as described in \citet{Boubert2021}.

\subsection{Gaussian Process Prior}

The free parameters $\beta_{j\kappa}$ are modelled as a function of apparent magnitude $G$ and colour $C = G-G_\mathrm{RP}$. This is performed in discrete colour-magnitude bins with a Gaussian Process prior placed on each dimension independently
\begin{equation}
    \{\beta_{j\kappa}\}_{mc}  \sim \mathcal{GP}\left(\{G\}_m, \{C\}_c\right)
\end{equation}
where we use an independent Squared Exponential kernel for each of apparent magnitude and colour,
\begin{align}
    K(G,C,G',C') = \sigma^2\exp\left(-\frac{(G-G')^2}{2l_m^2}\right)\exp\left( - \frac{(C-C')^2}{2l_c^2}\right),
\end{align}
with apparent magnitude and colour length-scales $l_m$ and $l_c$ and a variance $\sigma^2$, all of which are free parameters of that kernel.

The mean of the Gaussian Process is also a free parameter. In this work we set the mean to zero everywhere as $x=0$ corresponds to $q=0.5$ which is also the mean of a uniform prior reflecting the fact that we have no \textit{a priori} information on the sample selection probabilities. This is not to suggest that our prior is ``uninformative'', as we will see in the results the prior does inform the model in regions of parameter space lacking data, however, we will explain in Section~\ref{sec:discussion} that these regions are not of significant concern.

Some readers may be more used to our method being an example of ``regularised regression''. Whilst this is correct we apply our priors explicitly on the parameters of the model as opposed to having a broader regularization term. Therefore the model will always be discussed in the context of Bayesian statistics with a likelihood function on the data and priors on all parameters.

\section{Binomial Statistics}
\label{sec:binomial}

The selection function fit is based on the Binomial likelihood. For each bin, the number of sources in the sub-sample $k$ is assumed to be drawn from a Binomial distribution, given the number of sources which could have been selected from the source catalogue in the given bin $n$ and the selection probability $q$.

In this section we'll briefly revise the likelihood function and describe the Beta-Binomial expected value and p-value test which will be used for analysing the results.

\subsection{Binomial Likelihood}

The source catalogue and subset are counted in HEALPix-colour-apparent magnitude bins. The overall likelihood is given by
\begin{align}
    \mathcal{L} &= \prod_{i=p,m,c} \, \mathrm{Binomial}(k_i\,|\,n_i, q_i)\nonumber\\
    & =  \prod_{i=p,m,c} \, \binom{n_i}{k_i} q_i^{k_i}\,(1-q_i)^{n_i-k_i}
\end{align}
where $p,m,c$ is the HEALPix-magnitude-colour bin index and $q_i$ is the model probability at the bin center (Eq.~\ref{eq:x_model}). Since the Binomial coefficient is independent of the selection probability, this can be dropped out as a constant and the log likelihood simplifies to
\begin{align}
    \log\mathcal{L} \sim  \sum_{i=p,m,c} k_i\log(q_i)\, + (n_i-k_i)\log(1-q_i)
\end{align}

To optimize the likelihood function in terms of model parameters, $\beta_{j\kappa}$, we use the L-BFGS-B algorithm \citep{lbfgsb} implemented in \textsc{scipy}. The boundaries are placed at $[-50,50]$ for the unscaled parameters for which the prior distribution is a unit variance Gaussian. In other words, a parameter would have to be a $50$-sigma outlier from the prior to reach the optimization boundaries.

\subsection{Beta-Binomial Posterior}
\label{sec:pexp}

For each bin, we can independently estimate the posterior distribution of $q$ which will be useful when testing the results. This is not used to fit our model but instead to help understand the results.

Consider the stars in one bin to be like identical marbles in a bag. You get handed the marbles one by one and choose whether keep the marble or give it away. Eventually, of the original bag of $n$ marbles, you're left with $k$ in your hand. What is the expected probability of any marble being selected?

The Binomial distribution gives the likelihood of choosing $k$ marbles given the selection probability $q$,
\begin{align}
    \prob(k\,|\,n,q) &= \mathrm{Binomial}(k | n,q) \nonumber\\
    &= \binom{n}{k}\,q^k (1-q)^{n-k}.
\end{align}
To evaluate the posterior probability distribution of $q$, we apply Bayes theorem but first we need a prior. The Beta distribution is mathematically sensible as it is the conjugate prior of the Binomial distribution. Even more appealing, a Beta distribution with ${\alpha_0 = \beta_0 = 1}$ is equivalent to a uniform distribution, $ \mathrm{U}[0,1]$. The posterior probability distribution is
\begin{align}
    \prob(q\,|\,n,k) &= \frac{\mathrm{Binomial}(k | n,q)\,\mathrm{Beta}(q | \alpha_0, \beta_0)}{\prob(k\,|\,n)} \nonumber\\
    &= \mathrm{Beta}(q| k+\alpha_0, n-k+\beta_0).
\end{align}
This is the Beta-Binomial distribution with an expected $q$ of
\begin{align}
    \mathbb{E}[q] &= \frac{\alpha}{\alpha+\beta} \\
    &= \frac{k+\alpha_0}{n+\alpha_0+\beta_0}.
\label{eq:pexp}
\end{align}
For a uniform prior this equates to $(k+1)/(n+2)$. This formula is more commonly known as the ``rule of succession'' and was written down by Laplace more than two hundred years ago \citep{Laplace1921}.

This might not be quite what one expects. Naively $k/n$ is often used as the expected value of the selection probability given $k$ objects drawn from a sample of $n$. 

It is worth noting that the expected value is $1/2$ when $k=n=0$, which is the expected value of a uniform distribution. If there are no stars in a bin then we have no information to work with and the posterior reverts to the prior. This can be seen happening in the brighter bins in Figs.~\ref{fig:ast_fullhpx} and \ref{fig:ruwe_fullhpx_magonly}.

\subsection{p-value test}
\label{sec:pvalue}

We can test the veracity of a selection probability model using a p-value test (in this case we use a one-tailed p-value). This answers the question `Given the model, what is the probability that a measurement of an observable would not be larger than the observed value?'. For a Binomial distribution, the question is `What is the probability that less than $k$ sources out of $n$ are observed in this bin given the bin's selection probability?'. The Binomial p-value is given by
\begin{align}
    p_\mathrm{value} \sim \mathrm{U}\left[\sum_0^{k'=k-1} \mathrm{Binomial}(k'\,|\,n,q), 
    \sum_0^{k'=k} \mathrm{Binomial}(k'\,|\,n,q)\right].
\end{align}
which we explain in more detail in Appendix~\ref{app:binom}.
Note that the p-value is not a deterministic value but a random variable. For example, if we have no data in a bin ($k=n=0$) the p-value simplifies to $p_\mathrm{value} \sim \mathrm{U}\left[0,1\right]$. 

As in any one-tailed p-value test, if the model has successfully reproduced the data, the set of p-values for all data points will be uniformly distributed between 0 and 1. This test will be used with all of our inferred selection functions to check that they have accurately captured the information in the bins.

\section{Data}
\label{sec:data}

\begin{figure*}
  \centering
  \includegraphics[width=\textwidth]{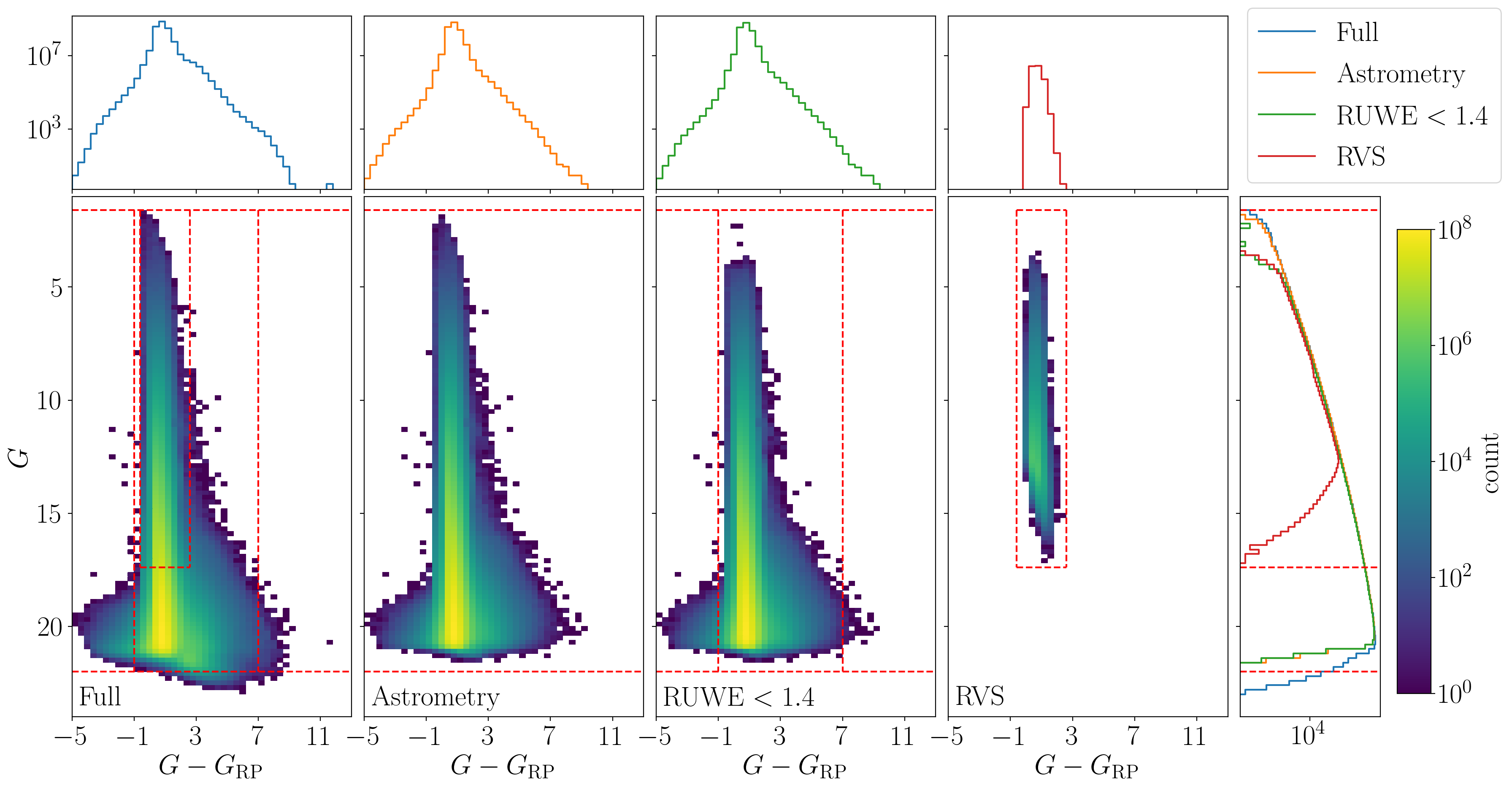}
  \caption[]{The number of \gaia EDR3 sources in each colour-magnitude bin is shown for the full sample (left) astrometry sample (middle left) astrometry with $\mathrm{RUWE}<1.4$ sample (middle right) and sample with DR2 radial velocity pubilshed in the EDR3 catalogue (right). The astrometry selection removes sources from the dim end whilst RVS is cut on $G_\mathrm{RVS}$ producing an extended drop-off in $G$ as can be sen from the histogram on the right. Red dashed lines show the region of parameter space used to fit the selection functions.}
   \label{fig:cm_count}
\end{figure*}

\begin{figure*}
  \centering
  \includegraphics[width=\textwidth]{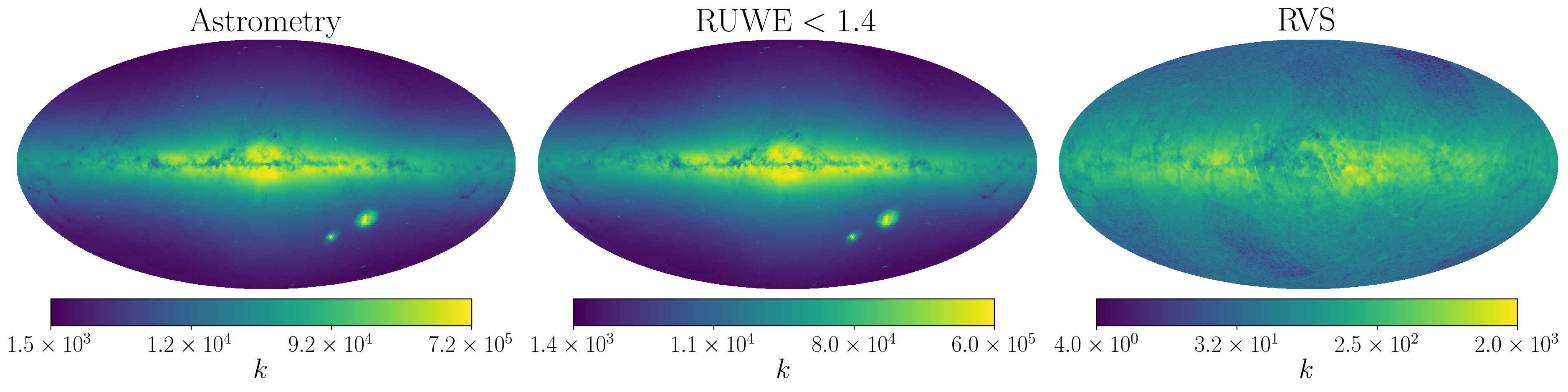}
  \caption[]{The counts of each sample in HEALPix level 6 bins across the sky summed over the colour-apparent magnitude range used for the selection function. The RUWE sample shows slightly reduced star counts in the highest density regions compared with the Astrometry and RVS is significantly more spatially uniform due to the bright magnitude limit biasing the spectrograph towards nearby sources which are more spatially uniform and because it is challenging for RVS to assign individual windows to sources in crowded regions. The RVS star counts also show features of the scanning law and vertical stripes in the East and squares in the Equatorial South deriving from the SDSS and Schmidt photographic plate contributions to the Initial Gaia Source List \citep{Rybizki2021RVS}.}
   \label{fig:k_sample}
\end{figure*}

\begin{figure*}
  \centering
  \includegraphics[width=\textwidth]{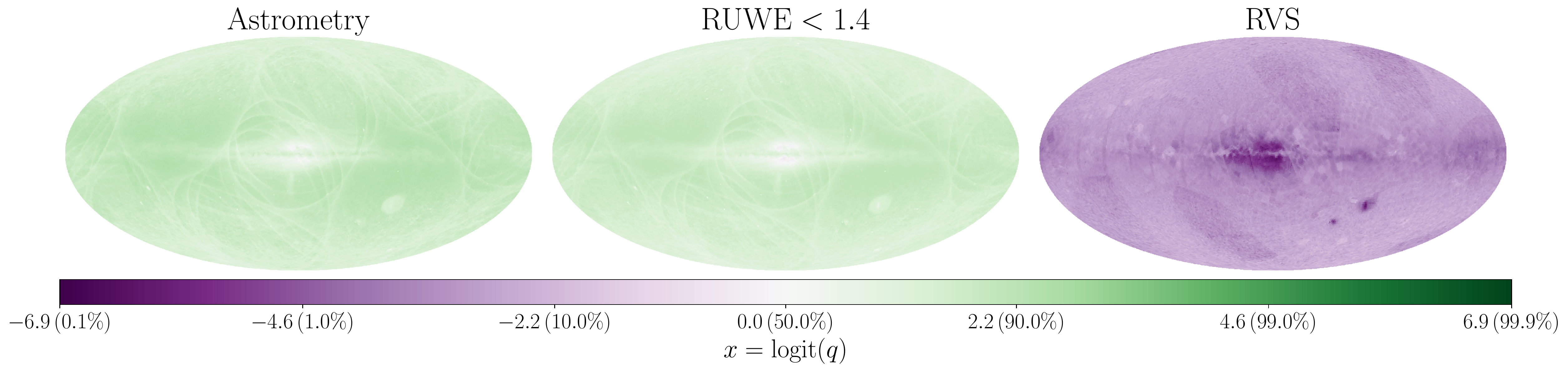}
  \caption[]{$q = (k+1)/(n+2)$ gives the expected binomial probability as explained in Section\ref{sec:pexp}. This is shown for each of the samples with $k$ and $n$ taken as the number of sources in each HEALPix level 6 bin summed over the colour and apparent magnitude range used for the selection function in the subsample and full samples respectively. In all three cases, the samples are limited relative to the source catalogue in crowded regions such as the bulge, LMC and SMC whilst there are also features of the scanning law which are apparent.}
   \label{fig:exp_x}
\end{figure*}

The selection function is estimated for three samples: Astrometry, RUWE and RVS. Here we'll provide a brief description of each sample and how it can be accessed from the \gaia archive.

\subsection{Astrometry}

The astrometry sample is the subset of \gaia EDR3 with published parallax and proper motions from the Astrometric Global Iterative Solution \citep[AGIS,][]{Lindegren2012}. There are three criteria for a source having published parallax and proper motion \citep{Lindegren2021}:
\begin{itemize}
    \item $N_\mathrm{VPU}\geq 9$
    \item $G_\mathrm{DR2}<21.0$
    \item $\sigma_\mathrm{5Dmax} < 1.2\mathrm{mas} \times \gamma(G)$.
\end{itemize}
The VPU (\textsc{visibility\_periods\_used}) cut depends on the number of observations a source has received and their distribution in time. This produces a strong dependence of the selection function on position on the sky. The apparent magnitude cut is performed on DR2 apparent magnitudes and all magnitudes have been recomputed in EDR3. Therefore the $G_\mathrm{DR2}$ cut doesn't produce a discontinuous change in the selection function with $G$ however it will still lead to a distinct drop-off towards the faint end. Finally, the cut on $\sigma_\mathrm{5Dmax}$ (\textsc{astrometric\_sigma5d\_max}) will depend on position on the sky and apparent magnitude, because both contribute to the expected astrometric uncertainty for any source as discussed in \citet{CoGIV}. The effects of these cuts will be empirically modelled in this work without directly considering the effects of each cut on their own. 

The distribution of sources as a function of colour and apparent magnitude is shown in Fig.~\ref{fig:cm_count} and as a function of position on the sky in Fig.~\ref{fig:k_sample}. The SQL query for accessing the data is as follows.
\begin{lstlisting}[language=SQL]
SELECT *
FROM gaiaedr3.gaia_source 
WHERE
    astrometric_params_solved>3
AND phot_g_mean_mag between 1.6 and 22
\end{lstlisting}

The method requires upper and lower limits for the final bounds of the magnitude bins. There are no sources in \gaia EDR3 with apparent magnitude brighter than $G=1.6$ and none with parallax or proper motion fainter than $G=22$ so we use these as our magnitude limits. We discuss the selection function outside these limits in Section~\ref{sec:discussion}. This results in a sample of 1,465,211,050 sources out of a total 1,806,195,366 in \gaia EDR3 with $1.6<G<22$.

\subsection{RUWE}


When using \gaia astrometry, various cuts are often placed on the sample to generate a subset with maximal information and minimal systematics. One such cut recommended by DPAC is $\mathrm{RUWE}<1.4$ \citep{Brown2021}. Renormalised Unit Weight Error (RUWE) is the reduced chi-square of the individual position measurements of a source given the source's astrometric solution renormalised by the 41\textsuperscript{st} percentile of the catalogue as a function of colour and apparent magnitude\footnote{RUWE is explained in more detail in the document ``Re-normalising the astrometric chi-square in Gaia DR2'' which is available from \url{https://www.cosmos.esa.int/web/gaia/public-dpac-documents}}. Sources with large $\mathrm{RUWE}$ have scatter between astrometric measurements which is poorly fit by the linear astrometry model. A common cause of this is  binary motion \citep{Lindegren2018, Belokurov2020, Penoyre2020} however this can also be generated by source contamination in crowded regions and possibly even astrometric microlensing \citep{McGill2020}. Whilst these sources can be astrophysically interesting, they introduce systematic errors in the AGIS pipeline and it is recommended to remove the more extreme cases in order to clean the sample.

There have been other recommended cuts for removing spurious astrometric solutions on \textsc{ipd\_gof\_harmonic\_amplitude} \citep{Fabricius2021} or \textsc{astrometric\_gof\_al} \citep{Lindegren2021} however $\mathrm{RUWE}<1.4$ is commonly used in the community so we keep to this single cut here.

Whilst an ideal telescope of \gaia's configuration would have a reflection symmetric PSF in the perpendicular along-scan and across-scan directions, the PSF of the \gaia spacecraft is slightly asymmetric on the CCD panel \citep{Rowell2021}, introducing chromatic behaviour to the inferred centroid location \citep[for full discussion see][Secton 2.3]{Lindegren2021}. This propagates through to the astrometric solution where the unit weight error has a colour dependence. RUWE is the renormalised unit weight error where the renormalisation is to the $41^\mathrm{st}$ percentile of the unit weight error as a function of colour and apparent magnitude\footnote{\url{http://www.rssd.esa.int/doc_fetch.php?id=3757412}}. As a result, the $41^\mathrm{st}$ percentile is achromatic however the spread of RUWE through the observed population will still have a residual colour dependence as will the fraction of sources with RUWE>1.4. Therefore colour dependence is an important aspect of the RUWE<1.4 selection function.


The problem with modelling RUWE as a function of colour is that we require all sources to have published colour. Approximately 88\% of sources in the \gaia EDR3 catalogue have published $G_\mathrm{RP}$. Therefore the selection function we will actually be modelling is ${\prob(\mathrm{RUWE}<1.4\,|\,\mathcal{S}_\gaia, \mathcal{S}_{G_\mathrm{RP}})}$. If the event that a source has ${\mathrm{RUWE}<1.4}$ is entirely independent of the the event that it has published $G_\mathrm{RP}$ this probability is the same as ${\prob(\mathrm{RUWE}<1.4\,|\,\mathcal{S}_\gaia)}$.

Within the apparent magnitude range $1.6<G<22$ there are $1.81$ billion sources in \gaia, $1.40$ billion of which have $\mathrm{RUWE}<1.4$, therefore $${\prob(\mathrm{RUWE}<1.4\,|\,\mathcal{S}_\gaia)\sim77.6\%}.$$ $1.55$ billion sources in \gaia with $1.6<G<22$ have published $G_\mathrm{RP}$, $1.29$ billion of which have $\mathrm{RUWE}<1.4$ such that $${\prob(\mathrm{RUWE}<1.4\,|\,\mathcal{S}_\gaia, \mathcal{S}_{G_\mathrm{RP}})\sim83.3\%}.$$ Therefore, whether a source has published $G_\mathrm{RP}$ affects the probability that the source will have $\mathrm{RUWE}<1.4$. 

\begin{figure}
  \centering
  \includegraphics[width=0.49\textwidth]{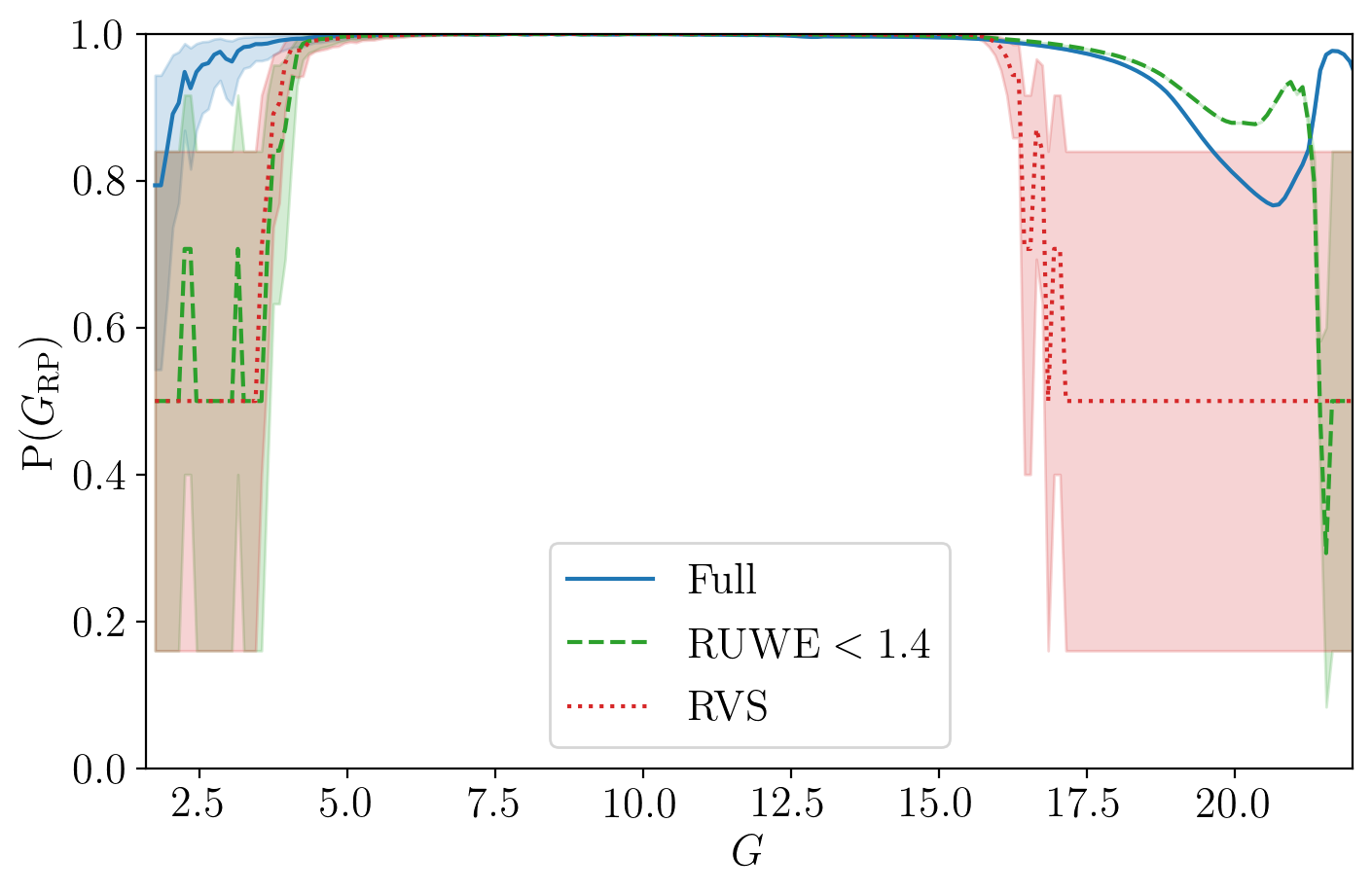}
  \caption[]{The probability of a source having $G_\mathrm{RP}$ is evaluated from the Beta-Binomial posterior distribution in Equation~\ref{eq:pexp}. For $G>15$ there are few enough sources in RVS that this isn't significantly different to the full sample probability however for $16\lesssim G\lesssim21$ the RUWE $G_\mathrm{RP}$ probability changes strongly from the full sample which could lead to significant biases in the RUWE colour selection function if not used with a correct $G_\mathrm{RP}$ selection function.}
   \label{fig:rp_prob} 
\end{figure}

We show this as a function of apparent magnitude in Fig.~\ref{fig:rp_prob} where we give the median and $16^\mathrm{th}-84^\mathrm{th}$ percentile ranges of the probability of a source having $G_\mathrm{RP}$ from the full sample or given the source satisfies the $\mathrm{RUWE}$ cut. For $16<G<21$, a source is significantly more likely to have $G_\mathrm{RP}$ if the source has published $\mathrm{RUWE}<1.4$.

Therefore we advise that the colour-dependent RUWE selection function should only be used in conjunction with the $G_\mathrm{RP}$ selection function
\begin{align}
    \prob(\mathcal{S}_\mathrm{RUWE}, \mathcal{S}_{G_\mathrm{RP}} \,|\, & \mathcal{S}_\gaia, \mathbf{y}) \\
    &= \prob(\mathcal{S}_\mathrm{RUWE} \,|\, \mathcal{S}_{G_\mathrm{RP}}, \mathcal{S}_\gaia, \mathbf{y})\,\cdot\,\prob(\mathcal{S}_{G_\mathrm{RP}}\,|\,\mathcal{S}_\gaia, \mathbf{y} ),\nonumber
    \label{eq:sruwe}
\end{align}
which is the probability of a source in EDR3 having both $\mathrm{RUWE}<1.4$ and published $G_\mathrm{RP}$.
Since the selection function for $G_\mathrm{RP}$ is not known, and for cases where one wishes to fit a model to \gaia data without colour dependence, we also fit a magnitude-only selection function to the RUWE data including all sources with published $G$ independent of whether $G_\mathrm{RP}$ was published. By evaluating this, we're implicitly marginalising over the colour distribution of sources
\begin{align}
    \prob(\mathcal{S}_\mathrm{RUWE} \,|\,& \mathcal{S}_\gaia, \mathbf{y})  \\
    &=\int \mathrm{d}C \,\prob(\mathcal{S}_\mathrm{RUWE}\,|\, \mathcal{S}_\gaia, \mathbf{y}, C) \cdot\prob(C \,|\,\mathcal{S}_\gaia, \mathbf{y})\nonumber
    \label{eq:sruweGonly} 
\end{align}
where $\prob(C\,|\,\mathcal{S}_\gaia, \mathbf{y})$ is the distribution of source colours in the \gaia source catalogue at the given position on the sky and apparent magnitude, $\mathbf{y}$.

\begin{figure}
  \centering
  \includegraphics[width=0.485\textwidth]{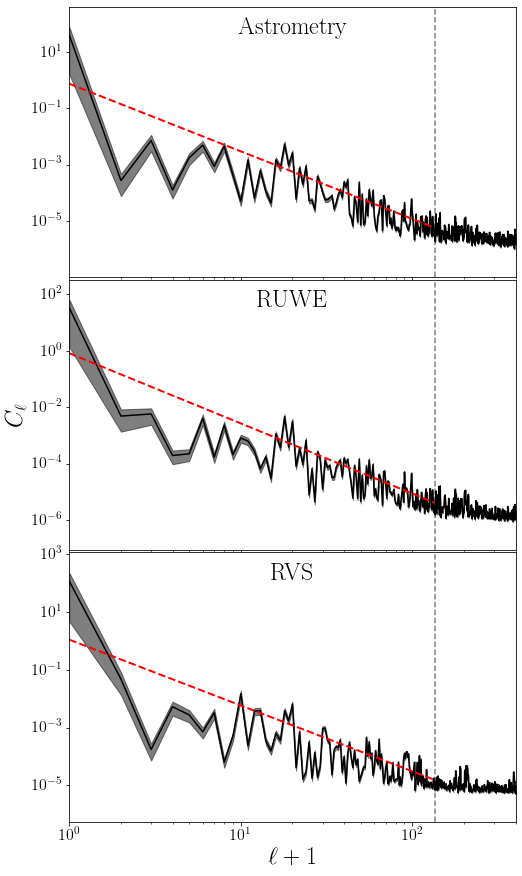}
  \caption[]{The power spectrum for each sample, evaluated using the spatial $(k+1)/(n+2)$ distribution from Fig.~\ref{fig:exp_x} shown by the black lines with grey $16^\mathrm{th}-84^\mathrm{th}$ percentile uncertainties, declines strongly with increasing $\ell$ in all samples. The red dashed line shows a power law fit for each sample out to $\ell=135$ which is the maximum $\ell$ set of spherical harmonics used to construct the needlets.}
   \label{fig:powerspec}
\end{figure}

\begin{figure*}
  \centering
  \includegraphics[width=\textwidth]{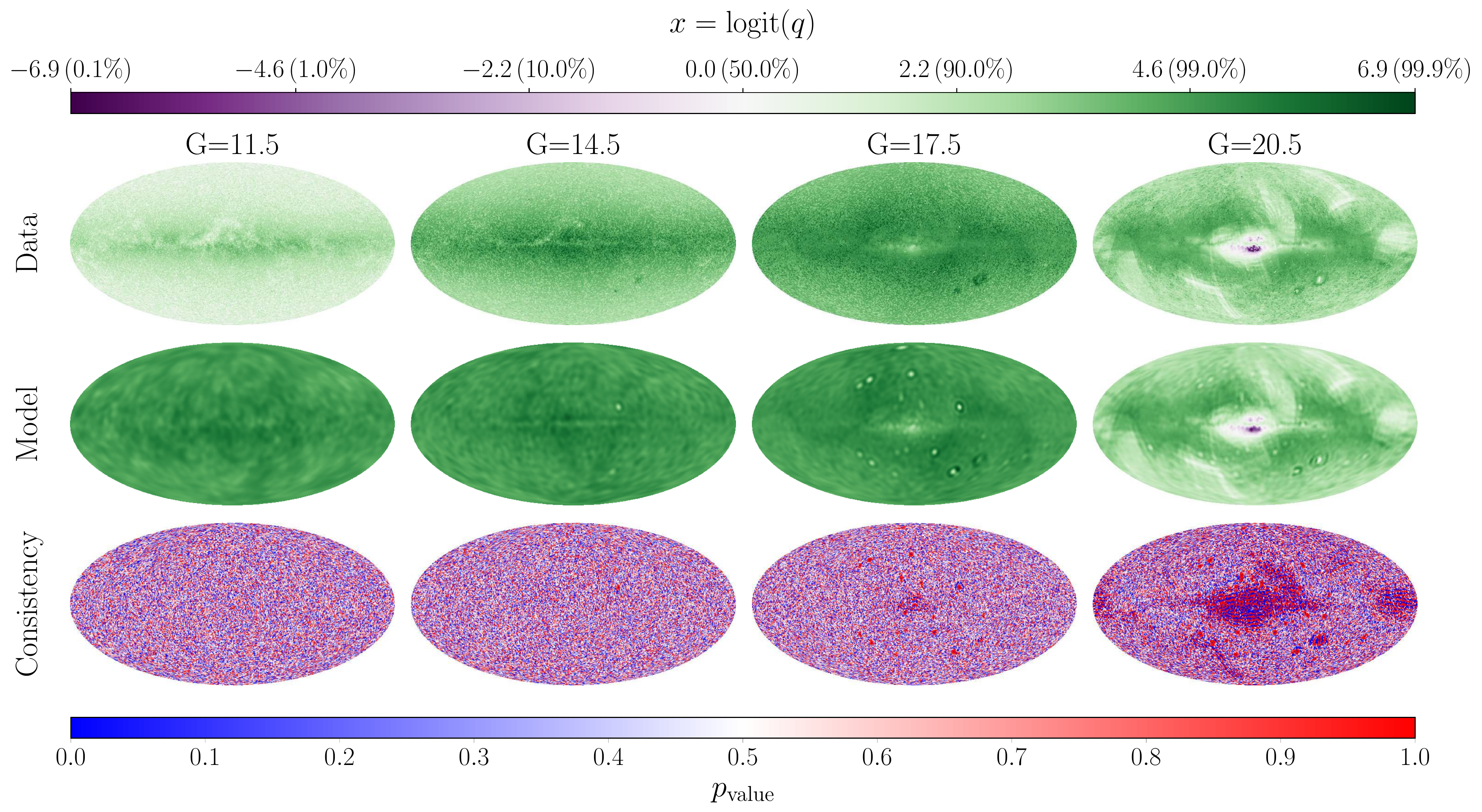}
  \caption[]{The results of the astrometry selection function fit (middle row) show a striking drop in high source density regions, not just in the disk and Magellanic clouds but also in the Milky Way globular clusters. There is also scanning structurex at the faint end which significantly reduces the selection function probability in under-scanned regions. The model shows strong agreement with the approximate expected distribution from the data (top row). This is demonstrated more precisely with the Binomial p-value tests which are dominated by random noise. In the faint bins, some structure arises due to lack of resolution in the model, as such our model should only be trusted to $\sim2$ degree scales.}
   \label{fig:ast_fullhpx}
\end{figure*}

\begin{figure}
  \centering
  \includegraphics[width=0.485\textwidth]{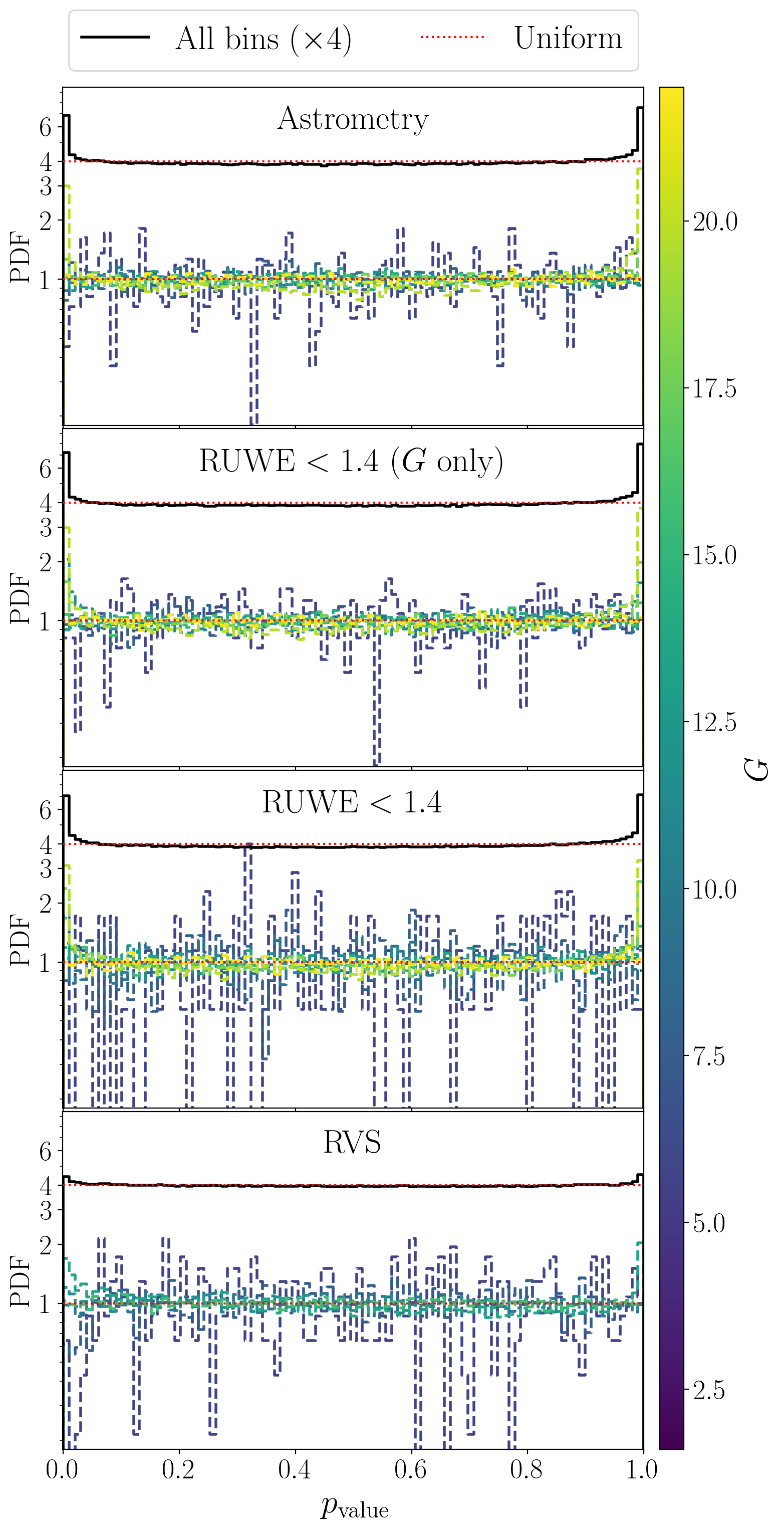}
  \caption[]{For all four models, the Binomial p-values of HEALPix-colour-magnitude bins are uniformly distributed at almost all magnitudes. This demonstrates that the model has fit the data exceptionally well. The minor exceptions are slight excesses of bins with p-values of 0 and 1 in the faintest bins of each sample which is caused by the lack of spatial resolution of the model which isn't able to pick up structure on scales under $2$ degrees. }
   \label{fig:pvalue_hist}
\end{figure}

The distributions of all \gaia sources and those with $\mathrm{RUWE}<1.4$ as a function of $G$ and $C$ are shown in the first and third panels of Fig.~\ref{fig:cm_count} respectively. Comparing with the \gaia Catalogue of Nearby Stars \citep{Smart2021} there are sources extending out to the blue ($G-G_\mathrm{RP}<-1$) and red ($G-G_\mathrm{RP}>3$) extremes of the colour distribution. The red sources will likely be extended sources or stars in crowded regions with high excess flux as mentioned in Section~\ref{sec:method}. The blue sources are very small in number and only appear at the faint end suggesting they are driven by photometric measurement errors for faint sources. We will only fit the colour-dependent selection for the range $-1<G-G_\mathrm{RP}<7$ which contains sources with well measured colours including those with high excess flux.

\begin{lstlisting}[language=SQL]
SELECT *
FROM gaiaedr3.gaia_source 
WHERE
    ruwe<1.4
AND phot_rp_n_obs>0
AND phot_g_mean_mag BETWEEN 1.6 AND 22
AND phot_g_mean_mag-phot_rp_mean_mag BETWEEN -1 AND 7
\end{lstlisting}

The RUWE sample has 1,292,152,210 sources out of a total 1,550,860,236 in the \gaia EDR3 source catalogue with colours and apparent magnitudes in the given range. 

For the colour-independent selection function, the RUWE sample is composed of 1,400,803,102 sources out of 1,806,195,366 in the same magnitude range in the \gaia source catalogue which can be retrieved with the following query.

\begin{lstlisting}[language=SQL]
SELECT *
FROM gaiaedr3.gaia_source 
WHERE
    ruwe<1.4
AND phot_g_mean_mag BETWEEN 1.6 AND 22
\end{lstlisting}
The source counts are shown as a function of position on the sky in Fig.~\ref{fig:k_sample} and predominantly trace the Milky Way stellar distribution.

\subsection{RVS}

The \gaia radial velocity sample is incredibly important to dynamical studies of the Milky Way \citep[e.g. ][]{Nitschai2020}. Radial velocities are measured by the \gaia satellite's on-board spectrograph using the calcium triplet. Radial velocities have not yet been released for \gaia DR3, however the DR2 sample of 7 million RVS stars is still by far the largest sample of stellar radial velocities available from any single observatory. The selection criteria used to produce the DR2 sample is given in \citet{Katz2018}. These are some of the main cuts:
\begin{itemize}
    \item $G^{\mathrm{ext}}_\mathrm{RVS}<12$ or $G^{\mathrm{int}}_\mathrm{RVS}<14$
    \item $3550<T_\mathrm{eff}<6900$
    \item No double line spectroscopic binaries or emission line stars.
\end{itemize}
$G^{\mathrm{ext}}_\mathrm{RVS}$ is the apparent magnitude estimated through photometric transformations of observations from ground-based observatories in the Initial Gaia Source List \citep[IGSL, ][]{Smart2014}. $G^{\mathrm{int}}_\mathrm{RVS}$ is directly estimated from the \gaia spectroscopy data in the radial velocity pipeline. In either case, $G_\mathrm{RP}$ provides a better approximation to $G_\mathrm{RVS}$ than $G$. However since the \gaia source catalogue selection function is modelled as a function of $G$, we keep to that here in the interests of simplicity and usability.

Selection on IGSL-measured $G_\mathrm{RVS}$ produces substantial structure across the sky as shown in the right-hand panel of Fig.~\ref{fig:exp_x} with vertical lines in the East due to SDSS \citep{Strauss2002} IGSL sources and a grid pattern around the South equatorial pole from the Guide Star Catalogue \citep{Lasker2008}. This is demonstrated and discussed in more detail in \citet{Rybizki2021RVS}.

Motivated by \citet{Boubert2019}, additional analysis and cleaning of the DR2 sample took place for the RVS sources published with EDR3 \citep{Seabroke2021}. 3,876 sources with incorrect radial velocities due to nearby neighbours were removed and 10,924 could not be successfully crossmatched with any source in \gaia EDR3. We also apply the cut $\textsc{rv\_nb\_transits}\geq 4$ recommended by \citet{Boubert2019} to clean out spurious radial velocity measurements.

The distribution of RVS sources as a function of apparent magnitude and $G-G_\mathrm{RP}$ colour is shown in the fourth panel of Fig.~\ref{fig:cm_count}. The RVS sample occupies a very narrow colour range and is limited to the bright end of the \gaia magnitude range. Only the range of colour and apparent magnitude containing radial velocity sources is used for the RVS sample, namely $-0.6<G-G_\mathrm{RP}<2.6$ and $1.6<G<17.4$, where we have applied the same bright-end cut as in the astrometry sample. 

Once again, we are faced with the same issue as in the RUWE selection where not all RVS sources have published $G_\mathrm{RP}$. However the RVS selection function is only non-zero at brighter magnitudes where $G_\mathrm{RP}$ is much more complete. For $1.6<G<17.4$, $6.2$ million out of $206$ million sources, or $3.00\%$, have published DR2 radial velocity which rises to $3.04\%$ of sources with published $G_\mathrm{RP}$. The RVS selection function has a much weaker dependence on $G_\mathrm{RP}$ selection than is the case for RUWE. By using the colour-dependent RVS selection function without accounting for the $G_\mathrm{RP}$ selection probability, a $\sim1\%$ systematic uncertainty would be introduced to the results. This is also shown in Fig.~\ref{fig:rp_prob} where a dotted blue line and shaded blue regions show that the $G_\mathrm{RP}$ probability is very high out to $G\sim16$ at which point we start to run out of RVS sources so the uncertainties become large.

\begin{lstlisting}[language=SQL]
SELECT *
FROM gaiaedr3.gaia_source 
WHERE
    dr2_rv_nb_transits>=4
AND phot_rp_n_obs>0
AND phot_g_mean_mag BETWEEN 1.6 AND 17.4
AND phot_g_mean_mag-phot_rp_mean_mag BETWEEN -0.6 AND 2.6
\end{lstlisting}
The RVS sample within the given colour-apparent magnitude range contains 6,186,950 out of a total 203,513,110 objects in the source catalogue within the same colour-magnitude range with published $G_\mathrm{RP}$.

For RVS, the distribution on the sky (shown in the right panel of Fig.~\ref{fig:k_sample}) is no longer solely dominated by the Milky Way source distribution. Aspects of the \gaia scanning law and features of the IGSL are visible in the on-sky distribution.


\subsection{Power Spectrum}
\label{sec:powerspec}



The Gaussian Process prior requires a dispersion parameter $\sigma$ which needs to be well chosen for the problem. The expected variance will depend on the Needlet scale.

We start from the expected selection probability across the sky, $\mathbb{E}[q] = \frac{\alpha}{\alpha+\beta} = \frac{k+1}{n+2}$ (see Section~\ref{sec:pexp}). The distribution of $\mathrm{logit}\left(\frac{k+1}{n+2}\right)$ is plotted across the sky for each of the samples in Fig.\ref{fig:exp_x}. A spherical harmonic model can be directly estimated for the distribution from the HEALPix values using
\begin{equation}
    \hat{a}_{\ell m} \sim \frac{4\pi}{N_\mathrm{pix}} \sum_{p=0}^{N_\mathrm{pix}-1} Y_{\ell m}^*(l_p, b_p) x_p
\end{equation}
where $x_p$ are the pixel values and $l_p,b_p$ are the Galactic longitude and latitude of the pixels. The power spectrum is then estimated by taking the square mean of mode amplitudes
\begin{equation}
    \hat{C_{\ell}} = \frac{1}{2\ell+1}\sum_{m=-\ell}^{\ell} |\hat{a}_{\ell m}^2|.
\end{equation}
This is done for each sample with the power-spectra shown by the black lines in Fig.~\ref{fig:powerspec}.

The power spectrum gives the expected variance of spherical harmonic coefficients. We model the power spectra with a single power-law distribution
\begin{equation}
    C_{\ell} = A(\ell+1)^\gamma
    \label{eq:plmodel}
\end{equation}
where $\gamma$ describes how the power decays for smaller scales. 

The sum of square spherical harmonic coefficients renormalised by their uncertainty is chi-square distributed with $2\ell+1$ degrees of freedom.
\begin{equation}
   \sum_{m=-\ell}^{\ell} \frac{|\hat{a}_{\ell m}^2|}{\sigma_\ell^2} \sim \chi^2(2\ell+1)
\end{equation}
where $\sigma_\ell^2 = C_\ell$ is the expected coefficient variance. Therefore we can use this to derive the likelihood of the given power spectrum model
\begin{equation}
    \prob(\hat{C_\ell}\,|\,C_\ell)  \propto z^{\frac{2\ell+1}{2}-1} \exp\left(-\frac{z}{2}\right)
\end{equation}
where $z=\frac{(2\ell+1)\hat{C_\ell}}{C_\ell}$. The $16^\mathrm{th}-84^\mathrm{th}$ percentiles of this distribution provide the shaded regions in Fig.~\ref{fig:powerspec}.

This is maximized with respect to $A$ and $\gamma$ in Eq.~\ref{eq:plmodel} to determine the best fit parameters using gradient descent with the Newton Conjugate Gradient method as implemented in \textsc{scipy}. We only use data with $\ell<135$ as this is the smallest scale spherical harmonic used in the Needlets and corresponds to a scale length on the sky $\sim 1$ degree which is the approximate pixel size of the data we will be using with HEALPix $\mathrm{nside}=64$. The best fit power law profiles are shown by the red dashed lines in Fig.~\ref{fig:powerspec} for the Astrometry, RUWE (magnitude-only) and RVS samples respectively with parameter values given in Table~\ref{tab:samples}.

Given the power spectrum of spherical harmonics, we want to work out what this implies for the variance of Needlet coefficients. From Appendix B of \citet{Boubert2021}, the variance of the Needlet coefficients as a function of the power spectrum is given by
\begin{equation}
    \langle |\beta_{j\kappa}^2|\rangle = \lambda_{j\kappa} \sum_{\ell=0}^{\ell_\mathrm{max}} b_l^2(j) C_\ell \frac{(2\ell+1)}{4\pi}
\end{equation}
where $C_\ell$ is taken from Eq.~\ref{eq:plmodel} using the best fit parameters and the normalisation constant is the area per pixel, $\lambda_{j\kappa}=\frac{4\pi}{N_j}$.

The reader should be concerned that we have done something uncomfortably non-Bayesian. We've used the data to determine the appropriate prior for our model. There are two reasons why we can get away with this. Firstly, we've used the data aggregated over colour-apparent magnitude space and only used this to estimate two parameters of a simple power spectrum such that the vast amount of information is hidden from our prior model decision. Secondly, and more importantly, in this work we're not attempting to infer a posterior distribution for our model. Using data to infer the prior would lead to an underestimation of posterior uncertainties as data has implicitly been double counted, however, we are only inferring the best fit selection function model.

The reason we've used the power spectra as described is that it makes the optimization significantly more computationally efficient as we're using an informed start point.

\renewcommand{\arraystretch}{1.5}
\begin{table*}
\begin{tabular}{c c c c c c} 
 \hline
  & & Astrometry & $\mathrm{RUWE}<1.4$ ($G$ only) & $\mathrm{RUWE}<1.4$  & RVS \\ [0.5ex] \hline\hline
Sample & $\sum\limits_{i=p,m,c} n_i$ & 1,806,195,366 & 1,806,195,366 & 1,550,860,236 & 203,513,110 \\
& $\sum\limits_{i=p,m,c} k_i$ & 1,465,211,050 & 1,400,803,102 & 1,292,152,210 & 6,186,950 \\
& $G$ range & [1.6,22] & [1.6,22] & [1.6,22] & [1.6,17.4] \\
& $G-G_\mathrm{RP}$ range & - & - & [-1,7] & [-0.6,2.6] \\
\hline
Bins & HEALPix nside & 64 & 64 & 64 & 64 \\
& $G$ bins & 0.2 & 0.2 & 0.2 & 0.2 \\
& $G-G_\mathrm{RP}$ bins & - & - & 2.0 & 0.4 \\
\hline
Power Spectrum & $\log A$ & -0.2937 & -0.6206 & -0.2032 & 0.0937 \\
& $\gamma$ & -2.3985 & -2.3552 & -2.4845 & -2.2761 \\
\hline
Model & $j_\mathrm{max}$ & 5 & 5 & 5 & 5 \\
& $l_G$ & 0.3 & 0.3 & 0.3 & 0.6 \\
& $l_{G-G_\mathrm{RP}}$ & - & - & 3.0 & 0.6 \\
& $B$ & 2.0 & 2.0 & 2.0 & 2.0 \\
\hline\hline 
\end{tabular}
\caption{For each \gaia sub-sample selection function, we provide the key parameters describing the data and model.}
\label{tab:samples}
\end{table*}

\section{Results}
\label{sec:results}


\begin{figure}
  \centering
  \includegraphics[width=0.485\textwidth]{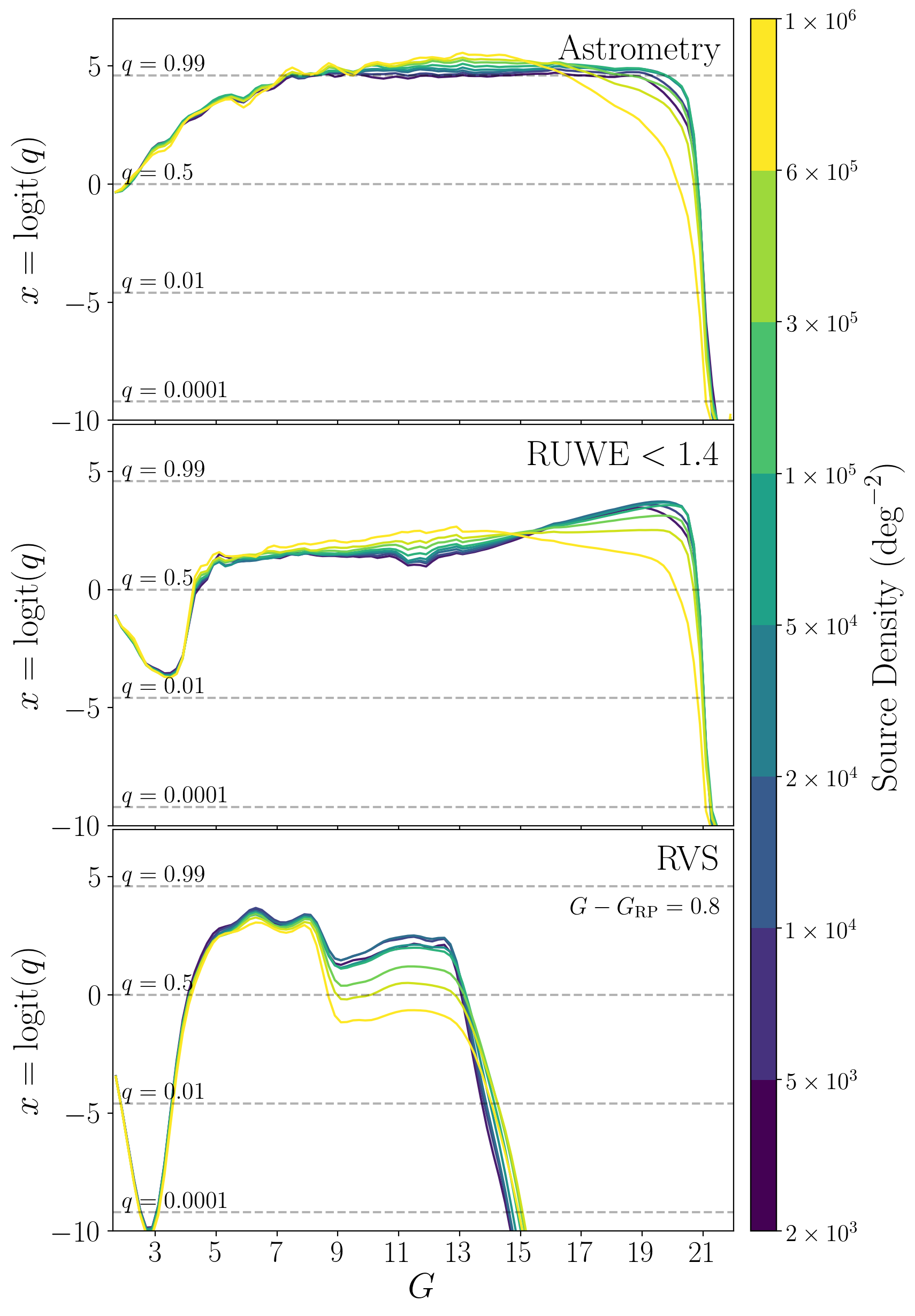}
  \caption[]{The selection function probability is strongly dependent on source density in all samples. By splitting pixels according to the source density in the full \gaia sample, we evaluate the selection function probability using the count-weighted average of all pixels within the given density bin. For the astrometry sample (top) the selection function drops at brighter magnitudes for higher source density regions. The $\mathrm{RUWE}<1.4$ colour-independent selection function (middle panel), shows a slightly different behaviour as much brighter sources are more likely to be cut out due to excess noise however the faint end shows similar behaviour to the astrometry. The RVS sample (bottom panel) at $G-G_\mathrm{RP}=0.5$ shows a more complicated pattern. At the bright end, as expected, the selection function is lower in high source density regions. However this flips at $G\sim13$. This is because RVS is limited in $G_\mathrm{RVS}\sim G_\mathrm{RP}$ and hence is more complete for redder sources as shown by Fig.~\ref{fig:x_cmd} which will largely be bulge fields due to dust extinction. In all samples, the model reverts to the prior mean ($x=0$) for $G\lesssim 3$ due to a lack of data in the source catalogue.}
   \label{fig:sd_mag}
\end{figure}

When estimating each of the selection functions we used a Needlet model with $j_\mathrm{max}=5$ and fit to data at HEALPix level 6 ($\mathrm{nside}=64$) resolution. This gives a total of $16,381$ Needlets with $\sim 2$ degree resolution fit to $49,152$ pixels. 
Details about the samples, binning schemes and model parameters are listed in Table~\ref{tab:samples}. A detailed discussion of the motivation behind the bin sizes and model parameters is provided in Section~\ref{sec:discussion}.

In this Section, we'll show the results of the fits as a function of sky position, apparent magnitude and colour. To test the veracity of the fits given the data, we'll use the Binomial p-value test explained in Section~\ref{sec:pvalue}.

\subsection{Astrometry}

The astrometry is fit in $0.2$ mag bins with a magnitude scale length of $0.3\;\mathrm{mag}$ - long enough that neighbouring bins are correlated, but short enough that the data can produce sharp changes in the model.

The astrometry selection function across the sky is shown in the middle row of Fig.~\ref{fig:ast_fullhpx}. The top row shows $(k+1)/(n+2)$ for the given magnitude bins, which, as discussed in Section~\ref{sec:pexp}, is the expected selection probability in the given magnitude bin independent of all others. High source density regions show much stronger selection limitations, particularly at the dimmer magnitudes. This is apparent through the white spots at $G=17.5$ which are centered on globular clusters such as $\omega$~Centauri and dwarf galaxies such as the LMC and SMC.

The bottom row of Fig.~\ref{fig:ast_fullhpx} shows the results of the p-value test discussed in Section~\ref{sec:pvalue}. At brighter magnitudes, p-values are distributed completely randomly demonstrating that the model has been successfully fit. Shifting to fainter magnitudes the selection function probability is reduced in regions of the sky with fewer scans. These `holes' in the scanning law are prominent due to the selection cuts on $N_\mathrm{vpu}$ and $\sigma_\mathrm{5Dmax}$ used for the EDR3 astrometry sample.

Structure also appears particularly around $\omega$~Centauri and the disk and LMC. This is where the selection function changes on scales smaller than the model can resolve. For example, the half-light radius of $\omega$~Centauri is $\sim4.8$ arc-minutes \citep{vandeVen2006} however the Needlets can only resolve structure on 2 degree scales. Therefore the reduction in selection probability due to the globular cluster is spread out over a wider area by the Needlet. At the core, in the pixel which contains the globular cluster, the selection probability is overestimated whilst being underestimated in any neighbouring pixels. There is a further-out halo of overestimated probability due to the structure of the Needlets which go negative before returning to zero \citep[see Fig.1 ][]{Boubert2021}.

We provide a histogram of p-values in the top panel of Fig.~\ref{fig:pvalue_hist} where we only include bins where $n>0$ as bins with $n=0$ have a uniformly distributed p-value independent of the model. The histogram for all bins is offset by a factor of 4 to make it clearer. For the vast majority of the data, we see well-behaved solutions with uniformly distributed p-values however, at the faint end, there are over-densities at $p_\mathrm{value}=0, 1$ where the resolution limitations become significant and we are under-fitting to the data.

The astrometry selection function as a function of magnitude is shown in the top panel of Fig.\ref{fig:sd_mag}. We group level 6 HEALPix pixels by number of sources in the pixel in the \gaia source catalogue with $1.6<G<22$. The selection function is the count-weighted mean of the selection functions in the given pixels.
For $G\lesssim 17$, the Astrometry sample is $\sim 99\%$ complete however this drops off quickly in high source density regions. Low density regions stay close to complete out to $G\sim 19$ before also falling rapidly. By $G\sim20$, less than $1\%$ of sources from the \gaia source catalogue are included in the astrometry sample.

For $G<1.6$ we are not able to say anything informative about the selection function probability as there is no data in the source catalogue here to use. For $G>22$ the selection function should be taken as zero.

For the astrometry sample, we have shown that the selection function model reproduces the observed data down to Needlet scales of $\sim 2$ degrees. At high latitudes, the astrometry sample can be complete out to $G\sim19$ and significantly drops at $G\sim21$ due to the cut placed on DR2 apparent magnitude. However the selection probability declines much brighter for crowded regions or where there are very few scans in EDR3.

\subsection{RUWE}

\begin{figure*}
  \centering
  \includegraphics[width=\textwidth]{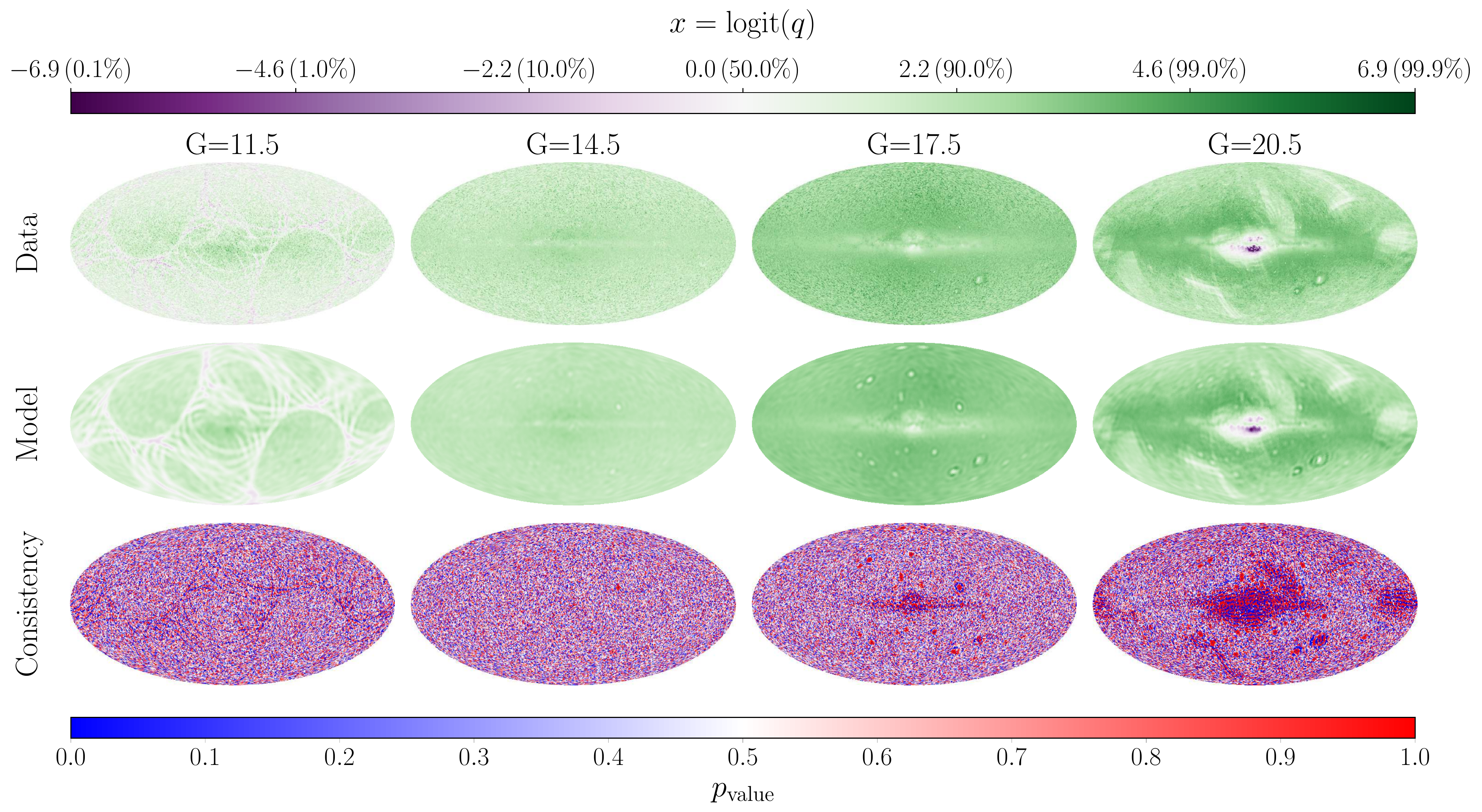}
  \caption[]{The $\mathrm{RUWE}<1.4$ selection function probability across the sky (middle row) shows high density region of sources at faint magnitudes due to crowding and scanning law features at bright magnitudes where the most precisely measured sources are more likely to have high RUWE due to excess source noise such as binary motion. Much of this structure is clearly visible in the $(k+1)/(n+2)$ plots (top row). The p-value test (bottom row) shows a good fit across most magnitudes and regions of the sky but shows that the model cannot fully resolve the scanning law at the bright end and source density structure at the faint end.}
   \label{fig:ruwe_fullhpx_magonly}
\end{figure*}

\begin{figure*}
  \centering
  \includegraphics[width=\textwidth]{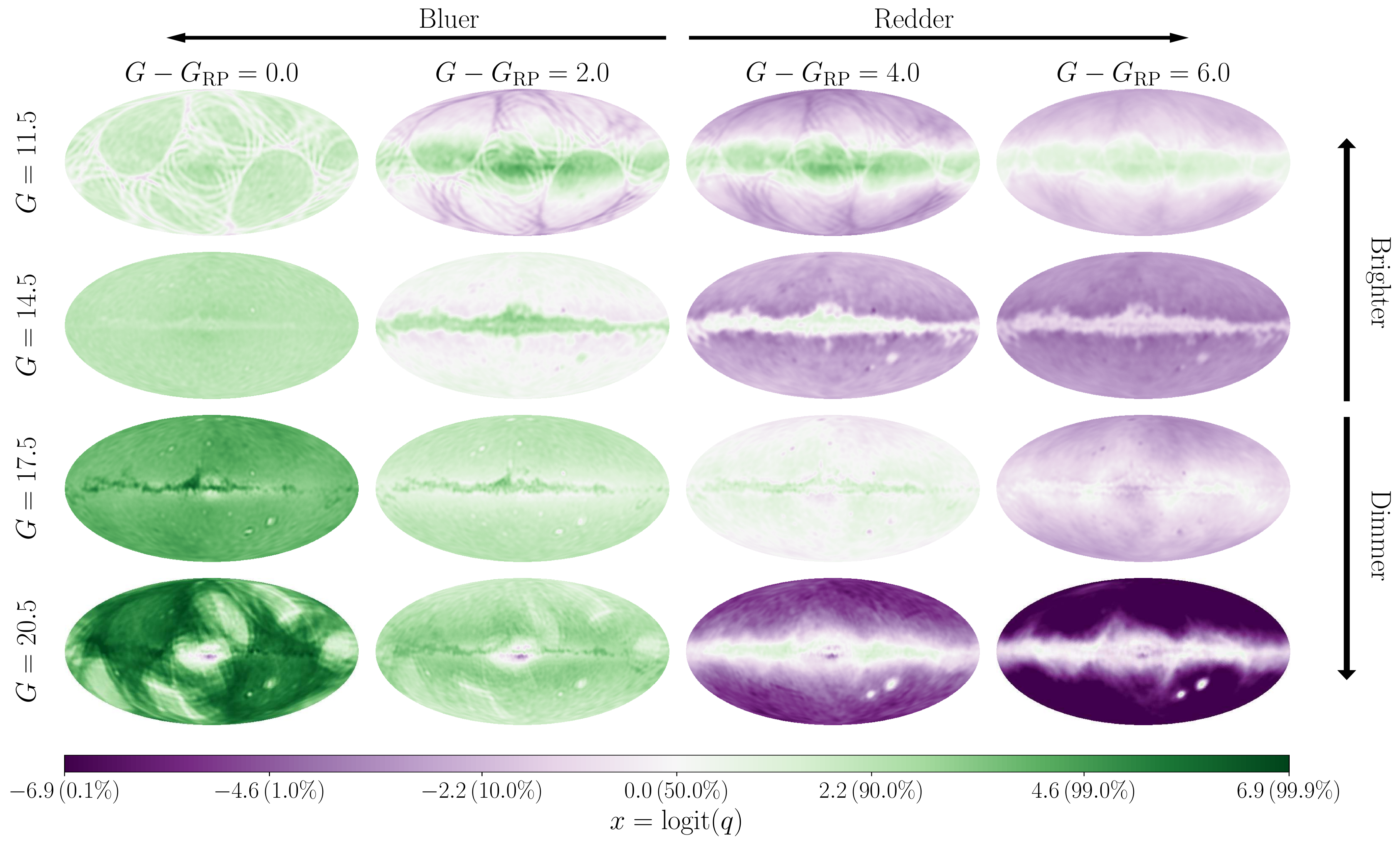}
  \caption[]{The RUWE selection function shows expected behaviour for blue faint bins (bottom left) where the probability is lower in high source density regions. However moving to redder and brighter magnitudes, this trend distinctly reverses with the disk showing significantly higher selection probability than high latitudes. An explanation would be that sources at high latitudes are typically nearer due to the sharp vertical drop off of the disk and so binary systems are more likely to produce a significantly large RUWE due to their large angular scale. The binary main sequence sits above and to the red side of the single star main sequence due to the combination of stellar fluxes which might explain why this reversal is more prominent at redder colours.}
   \label{fig:ruwe_xhpx}
\end{figure*}

\begin{figure*}
  \centering
  \includegraphics[width=\textwidth]{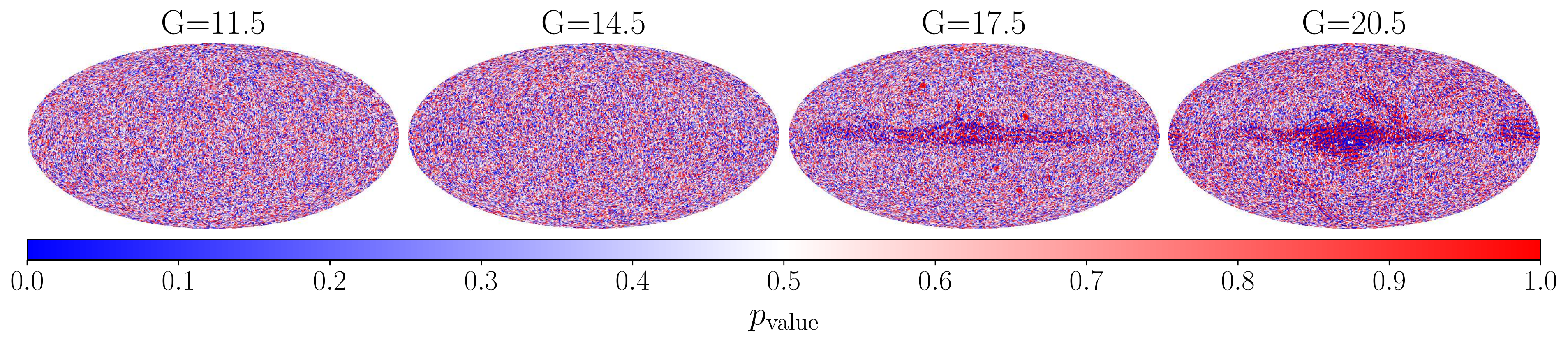}
  \caption[]{The binomial p-value for colour-magnitude bins of the $\mathrm{RUWE}<1.4$ sample selection function with $G-G_\mathrm{RP}\in [1.,3.]$ shows no residual structure at the bright end. At faint magnitudes the small scale structure of the disk shows up in waves as the model averages the variations out on $2$deg scales.}
   \label{fig:ruwe_pvalhpx}
\end{figure*}

\begin{figure*}
  \centering
  \includegraphics[width=\textwidth]{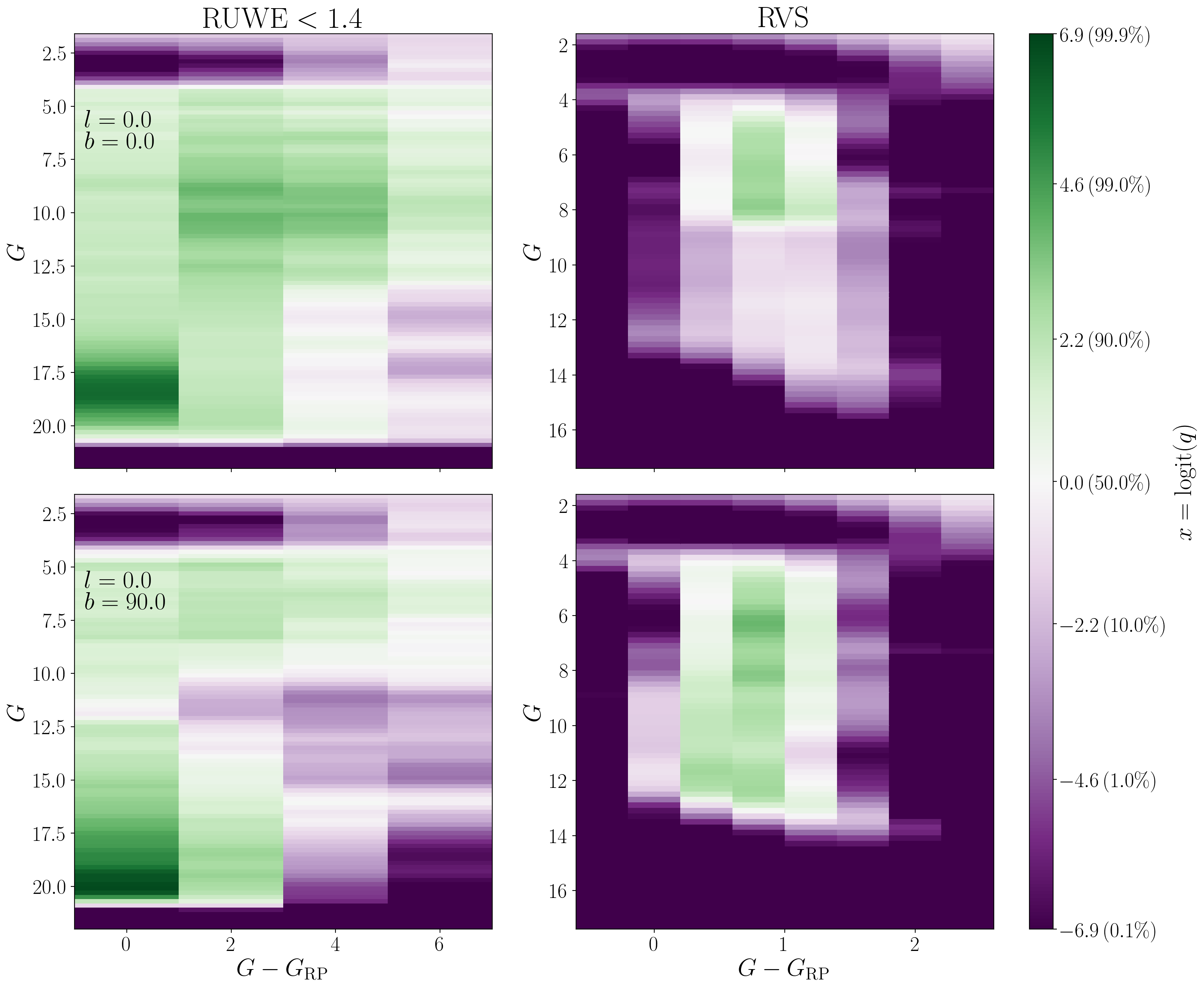}
  \caption[]{The RUWE selection function (left) is highest at faint blue magnitudes ($G\sim 18-20$, $G-G_\mathrm{RP}\sim 0$) and drops significantly at both the brightest and faintest ends for the crowded bulge field (top) and uncrowded Galactic pole field (bottom). There is also a strong gradient to lower selection probability for redder sources. This may be related to chromaticity of astrometric fits or relatively high completeness of $G_\mathrm{RP}$ at these colours.
  RVS has a narrow range of colours with significant selection probability which is likely the result of the $T_\mathrm{eff}$ selection used in the RVS pipeline \citep{Sartoretti2018}. At the faint end, the probability rapidly reduces but correlated with colour such that redder sources are observed to fainter magnitudes. This is due to the fact that RVS observes in the red part of the \gaia waveband and so the true RVS limit is closer to a $G_\mathrm{RP}$ cut. This produces the nice diagonal cut in $G$ vs $G-G_\mathrm{RP}$ corresponding to $G_\mathrm{RP}\sim12.5$.}
   \label{fig:x_cmd}
\end{figure*}

As discussed in Section~\ref{sec:data}, we evaluate the RUWE selection function in terms of sky-position and apparent magnitude against all sources in Gaia EDR3 and separately also as a function of ${C = G-G_\mathrm{RP}}$ for sources where $G_\mathrm{RP}$ is published. 

Both RUWE selection functions are fit to $0.2$mag bins in $G$ with a magnitude scale length of $0.3$. The colour dependent model includes four colour bins each $2$mag wide with a scale length of $3$mag which enables broad changes in the selection function probability as a function of $C$.

The results of the colour-independent model are shown in Fig.~\ref{fig:ruwe_fullhpx_magonly}. At the faint end the behaviour is similar to the astrometry selection function and the binomial p-value also shows that the fits are struggling to resolve features in crowded regions at faint magnitudes. For $G=17.5$ and brighter, however, the $\mathrm{RUWE}<1.4$ cut removes significant numbers of sources which have published astrometry. The selection probability is lower and for $G=11.5$ the scanning law is highlighted as heavily scanned regions are more likely to be removed by the cut on RUWE. 

The magnitude-dependent behaviour is shown more clearly in the middle panel of Fig.~\ref{fig:sd_mag} where the $\mathrm{RUWE}<1.4$ selection probability actually peaks near the faint end before declining and shows the same crowding dependence as the astrometry selection. However, the bright end has significantly reduced selection probability particularly in non-crowded regions. Our interpretation is that astrometric measurements have lower variance for bright sources and therefore sources with genuine intrinsic noise (such as binary systems) would have a more significant excess noise and be removed by the $\mathrm{RUWE}<1.4$ cut. For $G\lesssim 3$ the selection probability picks up again which is likely an artefact of the prior as there is very little data in the source catalogue at these magnitudes.

The colour-dependent selection function for $\mathrm{RUWE}<1.4$ across the sky for four colour and apparent magnitude bins is given in Fig.~\ref{fig:ruwe_xhpx}. Because RUWE is a measure of the astrometric error above the expected uncertainty, areas with high RUWE aren't necessarily those where \gaia performs worst but rather those where the satellite struggles unexpectedly. The selection function for $C=0$ at the dim end shows a similar structure to the astrometry selection suggesting that the astrometry sample cuts are the limiting factor in this region of colour space. As we shift towards redder sources the picture changes significantly. For redder sources, the low RUWE sample is more complete in the galactic plane and significantly less so in un-crowded regions. This inversion appears to be more extreme at brighter magnitudes.

This structure bares some similarities to Fig. 7 of \citet[][]{CoGIV} where the along-scan astrometric error is typically higher in crowded regions at bright magnitudes. A possible explanation is that the \gaia attitude error model either overestimates measurement uncertainties for bright sources in Galactic plane or underestimates those at high latitudes. This would result in higher RUWE for high latitudes and explain the observed selection pattern.

Fig.~\ref{fig:ruwe_pvalhpx} shows the p-values in four magnitude bins across the sky for the $C\in [1.0,3.0]$ colour bin. As in the astrometry selection, the bright bins have very successfully modelled the data however at the faint magnitudes, spatial resolution becomes important in the Galactic plane. Once again we see a rippling effect on scales which the Needlets aren't able to fully resolve. The second panel of Fig.~\ref{fig:pvalue_hist} gives the distribution of all p-values from which we can see that the resolution issues are confined to dim magnitudes and overall the model still fits the data extremely well.


The selection function probability for the RUWE sample across the CMD is shown in Fig.~\ref{fig:x_cmd}. For the highly crowded bulge region (top panel) the selection probability peaks at $G\sim 18$ in the blue and $G\sim 11$ in the red before declining towards fainter magnitude particularly for redder sources before cutting off sharply at $G\sim 20.5$. In a low-density field (bottom), there is a stronger bimodality with selection function peaks at $G\sim7$ and $G\sim 20$. This structure is likely connected to the different types of observations taken onboard \gaia with 1D windows for $G>13$ sources and 2D windows $G<13$ with window gating happening at still brighter magnitudes \citep{Evans2018}.

The overall trends with apparent magnitude for bluer sources are similar to the results from the colour-independent model shown in the middle panel of Fig~\ref{fig:sd_mag}. The cause of the drop-off in selection probability for faint red sources is unknown but may be related to chromaticity of the astrometric solution. An alternative explanation is that redder sources have lower $G_\mathrm{RP}$ and as such are more likely to have a measured $G_\mathrm{RP}$ in the source catalogue. This could push the selection function down at these colours. In this region of parameter space, one should make sure that the $G_\mathrm{RP}$ selection function is used in conjunction with $\mathrm{RUWE}<1.4$ as discussed in Section~\ref{sec:data}.

As in the magnitude-only model, the RUWE selection probability picks up for $G\lesssim 3$ due to lack of data in the source catalogue.

\subsection{RVS}

\begin{figure*}
  \centering
  \includegraphics[width=\textwidth]{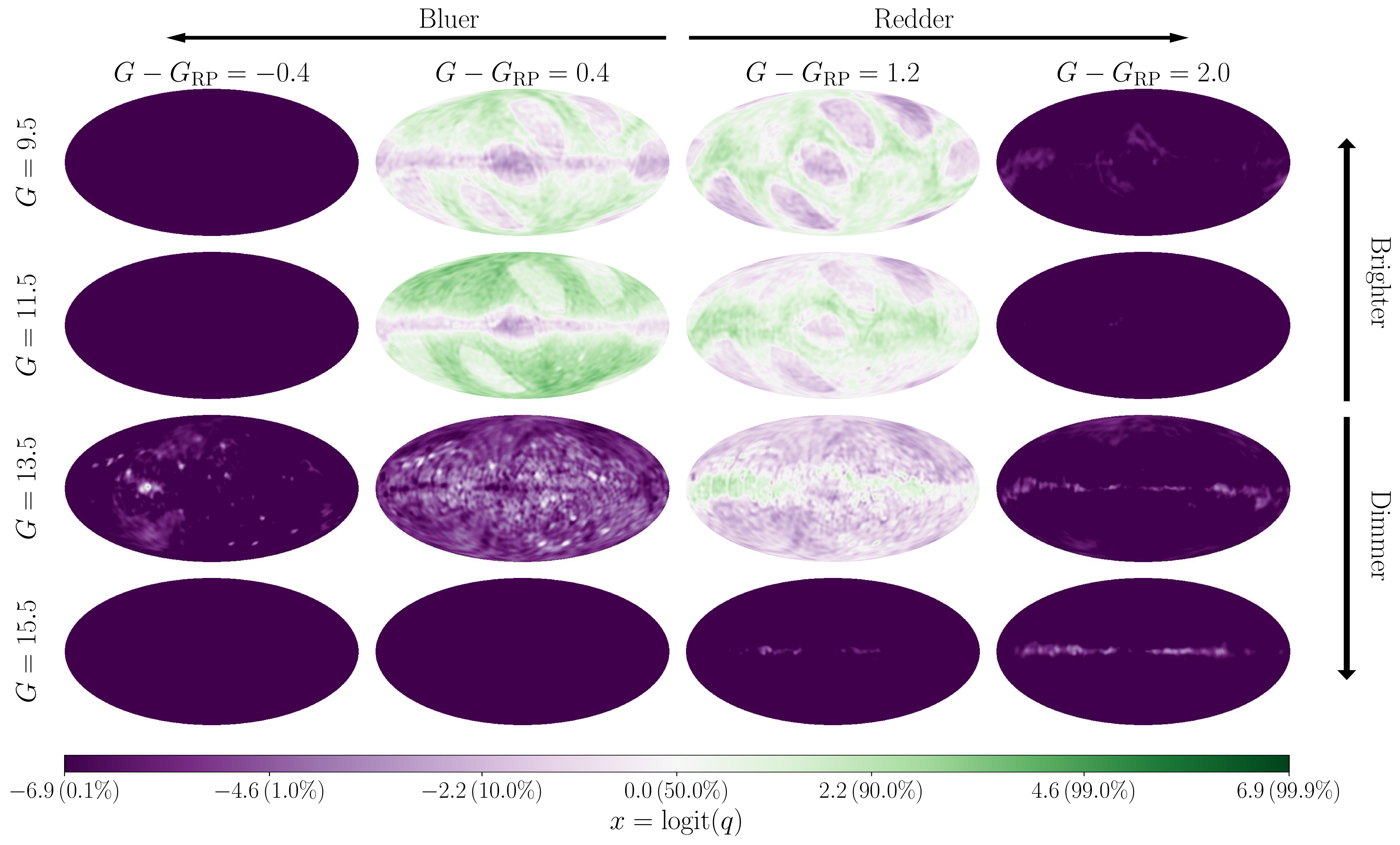}
  \caption[]{At brighter magnitudes (top rows) and bluer colours (left columns), the RVS selection probability is significantly higher in regions of the sky with more scans and lower source density. As we move dimmer and redder, this shifts to a higher selection probability in the Milky Way disk where dust extinction causes sources to be systematically redder and therefore have be relatively brighter in $G_\mathrm{RP}$ (see Fig.\ref{fig:x_cmd}).}
   \label{fig:rvs_xhpx}
\end{figure*}

\begin{figure*}
  \centering
  \includegraphics[width=\textwidth]{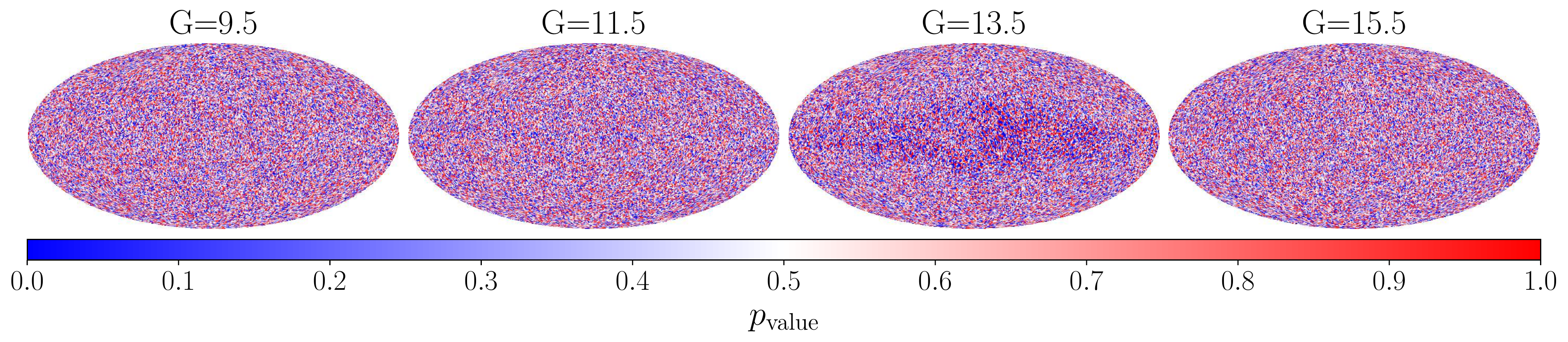}
  \caption[]{The p-value test for RVS bins, shown here for $G-G_\mathrm{RP}\in [0.6,1]$, is dominated by random noise with some small scale structure just visible at $G=13.5$. This shows that the model has provided an extremely good fit to the data.}
   \label{fig:rvs_pvalhpx}
\end{figure*}

The RVS sample is binned in $0.2$ mag bins in colour and $0.4$ mag bins in $G-G_\mathrm{RP}$. We use a scale length of $0.6$ in apparent magnitude and $1.2$ in colour. 
We opt for higher colour resolution than we did for $\mathrm{RUWE}$ because RVS selection will be closely related to $G_\mathrm{RP}$ which will generate strong ${C=G-G_\mathrm{RP}}$ dependence in the selection function.

The radial velocity sample is dependent on CCD observations from the on-board spectrograph which occupies four out of seven CCD rows on the \gaia focal plane. Depending on the $G_\mathrm{RVS}$ apparent magnitude evaluated for the source, a variety of window classes will be used as described in Section~7.1 of \citet{Cropper2018}. Sources with $G_\mathrm{RVS}<7$ (Class 0) received a full 2D windows whilst fainter sources only received 1D windows (Class 1 or 2). In crowded regions, Class 1 and 2 sources with overlapping windows would have their windows truncated in the region of overlap, often leading to non-rectangular windows. These observations were not used in DR2 data processing (as explained in \citealp{Sartoretti2018})  leading to 40\% of spectra being removed and a much higher percentage in crowded regions.

This process is manifested in the RVS selection function. The third panel in Fig.~\ref{fig:sd_mag} shows that the RVS selection function for $G\lesssim8$ (corresponding to approximately $G_\mathrm{RVS}\lesssim7$) is source density independent whereas sources in crowded regions are much less likely to be selected with $8\lesssim G \lesssim 12.5$.

For $G\gtrsim12.5$, this behaviour changes significantly. At dimmer and redder magnitudes, the RVS population traces the distribution of dust with high completeness in regions of the sky with significant extinction. We don't know the precise cause of this however there are some plausible explanations. The RVS catalogue filters out stars cooler than $T_\mathrm{eff}=3550$K which corresponds to $G-G_\mathrm{RP}\gtrsim 1.2$ \citep[see Fig.3][]{Andrae2018}. However, in regions with high dust extinction we will have hotter stars appearing with higher $G-G_\mathrm{RP}$ due to reddening. Therefore, for redder colours, the RVS cut on $T_\mathrm{eff}$ will be more strict where there is less extinction and produce a higher selection probability in dusty regions of the sky.


As discussed earlier, the RVS sample is selected on proxies for the apparent magnitude in the RVS waveband, $G_\mathrm{RVS}$. This is quite close to the $G_\mathrm{RP}$ waveband. A sharp cut in $G_\mathrm{RP}$ will lead to a correlated cut between $G$ and $G-G_\mathrm{RP}$. This is exactly what we see at the faint end of Fig.~\ref{fig:x_cmd}. For the bulge line of sight, the selection probability also drops off at much brighter magnitudes due to the crowding limits of the spectrograph. The narrow range of colour with non-zero selection probability is reflective of the RVS cuts on source effective temperature. The effect of the $G_\mathrm{RVS}$ cut is most distinctive in the third row, second column of Fig.~\ref{fig:rvs_xhpx} where the complex structure of the IGSL is very prominent.

The p-value test, shown across the sky in Fig.~\ref{fig:rvs_pvalhpx} for $G-G_\mathrm{RP}=0.8$, demonstrates that the model is correctly representing the data across most magnitudes. The third panel of Fig.~\ref{fig:pvalue_hist} also shows this with only one magnitude bin showing any poorly fit pixels with $G\sim14$. This may be the model struggling to reproduce sharp changes in the behaviour of the IGSL. 

As we saw for the RUWE model, the RVS selection function shoots up at $G\lesssim 3$ which can be seen in both Fig.~\ref{fig:sd_mag} and Fig.~\ref{fig:x_cmd} due to a lack of objects in the source catalogue. Users should be wary of this when applying the selection functions at these magnitudes.

\subsection{RVS and Astrometry or RUWE}

\begin{figure}
  \centering
  \includegraphics[width=0.485\textwidth]{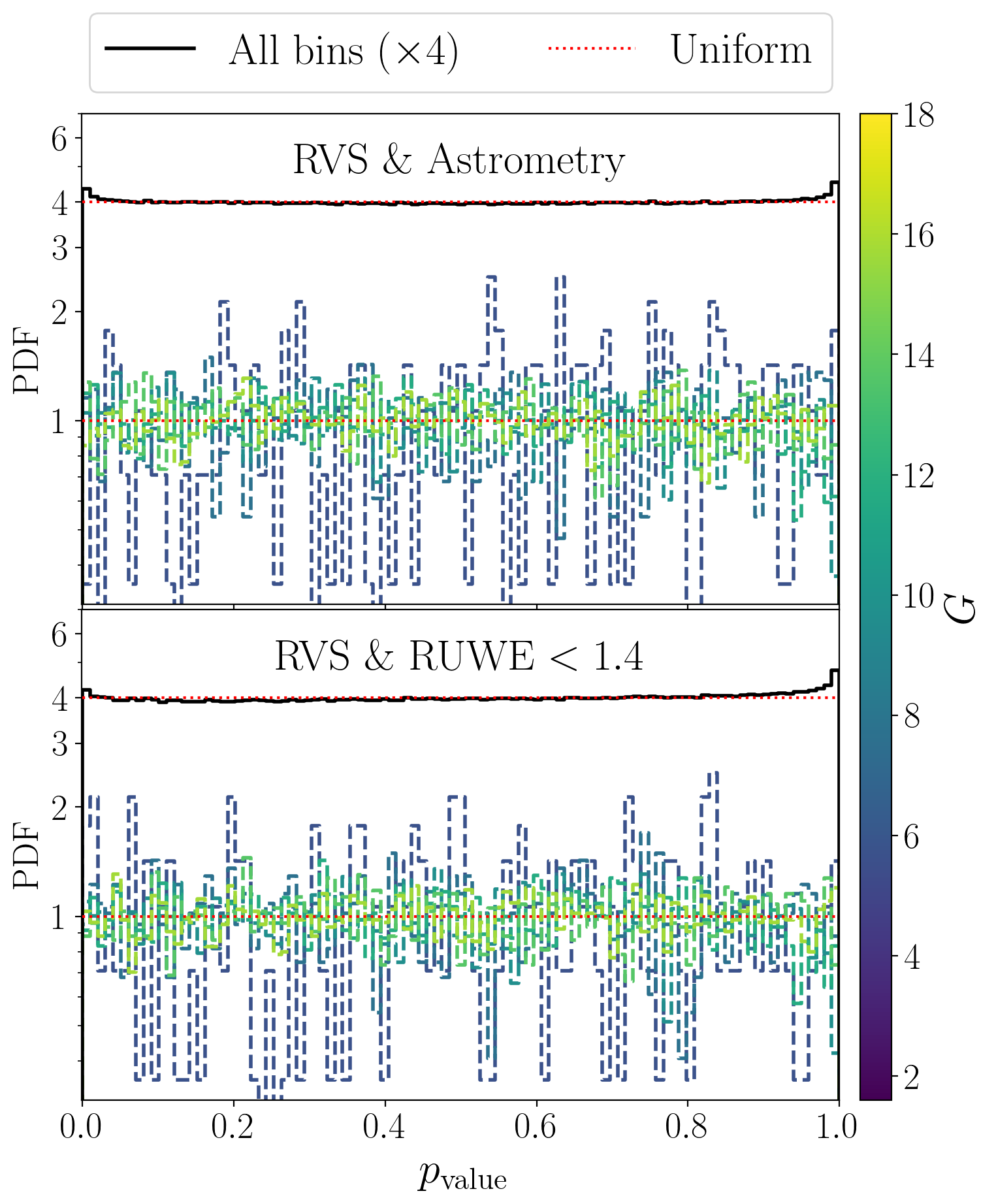}
  \caption[]{Distribution of p-values for samples with both RVS and Astrometry (top) and $\mathrm{RUWE}<1.4$ (bottom) compared with the model assuming independent samples is close to uniform suggesting that this is a reasonable assumption. In this case we've only included bins where $n>0$.}
   \label{fig:pvalue_comb}
\end{figure}

As well as individual subsets, various science cases will also require the intersection of subsets. For example, if one wanted full 6D phase space information for all objects in their sample, they'd want the subset of \gaia which has \textbf{both} published radial velocities and proper motion and parallax. 

The selection function for the intersection of two subsets is given by
\begin{align}
    \prob(&\mathcal{S}_{\mathrm{subset}1}, \mathcal{S}_{\mathrm{subset}2}\,|\,\mathcal{S}_\mathrm{source}, \mathbf{y}) \\
    & = \prob(\mathcal{S}_{\mathrm{subset}1}\,|\,\mathcal{S}_{\mathrm{subset}2}, \mathcal{S}_\mathrm{source}, \mathbf{y}) \,\cdot\,\prob(\mathcal{S}_{\mathrm{subset}2}\,|\,\mathcal{S}_\mathrm{source}, \mathbf{y}). \nonumber
\end{align}

The RUWE sample only contains sources with parallax and proper motion. Therefore ${\prob(\mathcal{S}_\mathrm{Astrometry}\,|\,\mathcal{S}_{\mathrm{RUWE}})=1}$ and the selection function for Astrometry and RUWE is equal to the RUWE selection function as expected.

If subset 1 and subset 2 are independent 
\begin{align}
    \prob(\mathcal{S}_{\mathrm{subset}1}\,|\,\mathcal{S}_{\mathrm{subset}2}, \mathcal{S}_\mathrm{source}, \mathbf{y}) 
     = \prob(\mathcal{S}_{\mathrm{subset}1}\,|\,\mathcal{S}_\mathrm{source}, \mathbf{y})
\end{align}
and we can take the selection function for the intersection to be the product of the two individual selection functions. 

We do this for the RVS sample where parallax and proper motion are provided and where $\mathrm{RUWE}<1.4$. There are 6,162,273 in \gaia EDR3 with both RVS and parallax and proper motion where 5,303,693 of these also have $\mathrm{RUWE}<1.4$. 

We take the product of the RVS position-colour-magnitude selection function defined in the previous section with the colour-independent selection functions for Astrometry
\begin{align}
    \prob(\mathcal{S}_\mathrm{RVS}, &\mathcal{S}_\mathrm{Astrometry}\,|\,\mathcal{S}_\gaia, \mathbf{y}) \\
    & = \prob(\mathcal{S}_\mathrm{RVS}\,|\,\mathcal{S}_\gaia, \mathbf{y}) \,\cdot\,\prob(\mathcal{S}_\mathrm{Astrometry}\,|\,\mathcal{S}_\gaia, \mathbf{y}) \nonumber
\end{align}
and RUWE
\begin{align}
    \prob(\mathcal{S}_\mathrm{RVS}, &\mathcal{S}_\mathrm{RUWE}\,|\,\mathcal{S}_\gaia, \mathbf{y}) \\
    & = \prob(\mathcal{S}_\mathrm{RVS}\,|\,\mathcal{S}_\gaia, \mathbf{y}) \,\cdot\,\prob(\mathcal{S}_\mathrm{RUWE}\,|\,\mathcal{S}_\gaia, \mathbf{y}). \nonumber
\end{align}
We then compare this with the data. Fig.~\ref{fig:pvalue_comb} shows the distribution of p-values for this sample where we only include bins with $n>0$ as bins with $n=0$ give a uniformly distributed p-value independent of the model. There is on significant sign of underfitting so we conclude that it is reasonable in this case to assume that RVS and Astrometry or RUWE samples are independent and therefore use the product of probabilities.

\section{Discussion}
\label{sec:discussion}

For each selection function we have made decisions about the variables to use, binning schemes to aggregate the data and the values of free model hyperparameters. In this section we justify the choices made.

\subsection{Selection function variables}

For our selection variables we decided to use $l, b, G$ and $G-G_\mathrm{RP}$ where appropriate. As discussed in Section~\ref{sec:method}, the cuts used to generate the given \gaia subsets are primarily dependent on these variables. However, we have found that all selection functions are strongly dependent on source density in the \gaia source catalogue. Whilst this is indirectly modelled through the spatial variables, directly using a crowding variable could enable us to better model the selection function in small dense objects such as globular clusters.

We could not use crowding as a selection variable in this work because it is not a measurable. The number density of sources in the \gaia source catalogue provides some insight but this is not the true underlying density of sources as \gaia is not a complete catalogue. The detection probability of a source in \gaia is also likely to primarily depend on the number density of sources with the same apparent magnitude or brighter such that a source density variable would also need to be magnitude dependent. As a further complication, we should really be modelling the source density on the focal plane. This includes sources from both fields of view separated by a basic angle of 106.5 deg such that the source density is also a function of observation time mapped through the scanning law. Constructing selection functions as a function of crowding is a complicated problem worth solving for \gaia subsets but beyond the scope of this paper.

This means that users should be wary of applying our selection functions to synthesised catalogues in order to generate mock \gaia data. The spatial component of the selection function includes strong crowding dependence which reflect features of the Milky Way such as the bulge, LMC and globular clusters but which will be different for any other simulated galaxy.

\subsection{Binning schemes}

In principle we could evaluate the likelihood for every source at its exact coordinates. With 1.8 billion sources, just evaluating the spatial model at all source positions with $j_\mathrm{max}=5$ would involve $\sim10^{13}$ function evaluations. And this would need to be done for every iteration of the optimization. Instead we aggregate the data in pixels and colour-magnitude bins to make the problem computationally tractable. Our choices of bin scales for each selection function are shown in Table~\ref{tab:samples}.

We have used HEALPix level 6 ($\mathrm{nside}=64$) such that the data is resolved on sub-degree scales. The selection function can change on smaller scales than this, particularly in globular clusters, however, it is computationally prohibitive to resolve the cores of globular clusters many of which have scale radii of arcminutes or smaller \citep{Vasiliev2021}.

The widths of magnitude and colour bins are decided based on our expectation of how fast we believe the selection function will vary in these parameters. The selection function is heavily magnitude dependent so we resolve in magnitude as high as we can within computational limitations. The colour dependence of RUWE is due to the rescaling of the unit weight error to account for the chromaticity of the astrometry. Therefore we only include the wide 2 mag bins in colour to encapsulate any general trends. RVS has a more significant colour dependence due to cuts placed on effective temperature so we use narrower 0.4 mag bins for this.

\subsection{Model parameters}

The maximum level of needlets used, $j_\mathrm{max}$ is set to one lower than the level of the data binning. We cannot use the same level or higher as this would lead to more model parameters than data points. This level of needlets enables us to fit a model which varies on $2$~deg scales across the sky.

$B$ and $\nu$ determine the shape and width of the needlets. The relation between these parameters and the needlet shape is complicated but by setting $B=2,\nu=1$ the width of the needlets is similar to the distance between needlet centers at a given HEALPix level. This enables us to fit smooth models across the sky without significant inter-needlet fluctuations.

The magnitude and colour lengthscales of the Gaussian Process, $l_G$ and $l_{G-G_\mathrm{RP}}$, are set to $1.5\times$ the bin widths in almost all cases. This allows for correlations between neighbouring data points and results in a smooth model as a function of colour and apparent magnitude. The one exception is $l_G$ for RVS which we set to $0.6$ to match $l_{G-G_\mathrm{RP}}$. As discussed in Section~\ref{sec:data}, the RVS sample is selected on $G_\mathrm{RVS}$ which is a narrow band in the same magnitude range as $G_\mathrm{RP}$. Any features at fixed $G_\mathrm{RVS}$ will produce a correlated feature in $G$ vs $G-G_\mathrm{RP}$ as we saw in Fig.~\ref{fig:x_cmd}. We set the scale lengths such that the model has equal power to resolve in magnitude and colour.

The mean and variance of the Gaussian Process are set such that the prior is only significant when there is very little data to fit. This occurs for $G<3$ at which point the model reverts to the mean as is seen in Fig~\ref{fig:sd_mag}. When using the selection function, users concerned about extremely bright stars should consider truncating their models at $G=3$ as the selection functions are not constrained by data brighter than this.

\subsection{Magnitude and colour bounds}

The selection functions are only defined within apparent magnitude and colour ranges given in Table~\ref{tab:samples} so a choice needs to be made about how to extrapolate the selection functions.

In all cases there are sources in the EDR3 source catalogue beyond the faint limit but none in the subset. The selection probability should be treated as zero fainter than this. As discussed in the previous subsection, any models using our selection function should be truncated for $G<3$ where the prior dominates the model.

We have no \textit{a priori} knowledge about the exact colour dependence of the RUWE selection function. We recommend that bluer or redder than the colour range we have modelled users should take the value of the $G-G_\mathrm{RP}=[-1,1], [5,7]$ bins respectively.

The RVS sample drops for blue and red sources due to cuts on effective temperature so the selection function should be taken as zero for $G-G_\mathrm{RP}<-0.6$ or $>2.6$.

\section{Other samples}

In this paper we have produced selection functions for three scientifically important sub-samples of the \gaia mission. However there are many catalogues in the \gaia-verse which are used by the astronomy community for a wide range of exciting research. Here we list some which were considered but which, due to additional complexity, we were not able to evaluate with the current method.

\subsection{Variables}

Variable sources are incredibly important as standard candles allowing us to measure precise distances to sources and systems across the Milky Way and in external galaxies \citep{Muraveva2018, Riess2019}. This has enabled detailed maps of the distribution of stars throughout the Milky Way disk and halo \citep{Skowron2019, Iorio2019}.

\citet{Holl2018} describes the \gaia DR2 variable sample following the processing detailed in \citet{Eyer2017}. As discussed in \citet{Holl2018}, overall completeness was not an aim of the DR2 variables sample. Considering only the lower limits placed on number of successful FoV transits required for variable processing they estimate that the average completeness is $\sim80\%$ for the $\mathrm{FoV}\geq12$ pipeline and as low as $\sim51\%$ for $\mathrm{FoV}\geq20$. This varies heavily across the sky due to the scanning law and crowding effects. Therefore a selection function for this sample would be hugely valuable for models of the Milky Way.

There are several reasons why the method we have applied here would not be directly applicable to variable samples. Firstly we need to consider what question we're asking. We want to know e.g. ``Given that a source with $l,b,G$ is an RRLyrae, what is the probability that \gaia would publish this source classified as an RRLyrae?''. Written as a probability, this is
\begin{equation}
    \prob(\mathcal{S}_\mathrm{RRLyrae}\,|\,l, b, G, X_\mathrm{RRLyrae})
\end{equation}
where $X_\mathrm{RRLyrae}$ is the event that the source is an RRLyrae-type star. Were we to naively use the approach we've used in this paper for other samples, we would actually evaluate the probability that a source is selected \textbf{and} it is an RRLyrae
\begin{equation}
    \prob(\mathcal{S}_\mathrm{RRLyrae}, X_\mathrm{RRLyrae}\,|\,l, b, G).
\end{equation}
This is because we'd be comparing the population of classified RRLyraes against the full sample of \gaia sources the vast majority of which are not RRLyrae stars.

To model the selection function for variable samples, we would need to know which sources in \gaia \textbf{would have been classified} as a variable had they actually been one.

The second issue is in choosing the observables over which to define the selection function. The selection function for variable sources will be highly dependent on variability amplitude and period. \gaia observes stars with certain dominant frequencies due to the spin, precession and orbital dynamics of the satellite. This would make some variability periods harder to detect than others. The apparent $G$-band magnitude is also highly unreliable for variable sources \citep[see Section 5.4 ][]{Arenou2018}. Improved apparent magnitudes may be determined from the time series however this adds another layer of complexity when comparing against the full \gaia dataset.

Finally, there are also many individual variable samples, such as RRLyrae, Cepheids and Long Period Variables, which are each used for separate science aims. A full study of the selection functions of variable sources in \gaia should aim to construct selection functions for each population.

The variables selection function for the \gaia mission is a significant, complex project worthy of dedicated further study.

\subsection{Parallax SNR}

$1/\varpi$ is often used as a distance estimate for sources with \gaia astrometry. 
However, for sources with low parallax signal-to-noise (SNR) the distance uncertainty distribution can be highly asymmetric and the measured parallax is negative for many sources in \gaia EDR3 \citep[see ][ for full discussion]{BailerJones2015, Luri2018}. 

A common selection to avoid this is taking only sources with $\varpi/\sigma_\varpi>X$ where commonly used values for $X$ are 5 or 10 depending on the scientific aims. This choice significantly affects the selection function as a function of distance and as a result, the inferred distance to sources can be heavily biased \citep[see][for a detailed discussion]{Schonrich2017}.  

One solution to this was demonstrated by \citet{Schonrich2019} who infer a distance-dependent selection function for the \gaia DR2 RVS sample. The limitations of the method used here is that they must assume an a-priori model for the distribution of sources in the Milky Way. As a result, their selection function is not model independent. Their selection function is also sky-averaged for the entire sample and doesn't look to quantify any of the complex variation of the selection function on small spatial scales.

We could use the method from \citet{Boubert2021} directly on the $\varpi/\sigma_\varpi>5$ sample as a subset of the \gaia catalogue and evaluate the selection function as a function of $G, G-G_\mathrm{RP}, l, b$. Indeed, $\sigma_\varpi$ is strong function of position on the sky due to the scanning law and a function of $G$ due to CCD photon noise \citep{CoGIV}. 

However the selection is also directly dependent on $\varpi$ so any selection function for a parallax SNR selected sample would need to feature measured $\varpi$ or observables which are directly dependent on it. For this reason, we do not attempt to model the parallax SNR selection function here.

\section{Access and use the selection functions}
\label{sec:access}

The selection functions are accessibly through the GitHub repository \textsc{selectionfunctions} (\url{https://github.com/gaiaverse/selectionfunctions}). Here we provide a brief example of how to load the EDR3 source catalogue and RVS subsample selection functions to estimate the probability that S5-HVS1 \citep{Koposov2020} could have been observed in \gaia EDR3 with radial velocity measure from DR2. The inferred probability is $\sim 3\times10^{-9}$ and indeed S5-HVS1 does not have published radial velocity.

\begin{lstlisting}[language=Python]
import selectionfunctions.cog_v as CoGV
import selectionfunctions.cog_ii as CoGII
from selectionfunctions.source import Source

# Load DR3 selection function
dr3_sf = CoGII.dr3_sf(version='modelAB',crowding=False)

# Load RVS selection function
rvs_sf = CoGV.subset_sf(map_fname='rvs_cogv.h5')

# Introduce source
s5_hvs1 = Source('22h54m51.68s',
           '-51d11m44.19s',
           photometry={'gaia_g':16.02,
                       'gaia_bp_gaia_rp':-0.008},
           frame='icrs')
           
# Selection probability
print(dr3_sf(s5_hvs1) * rvs_sf(s5_hvs1))

\end{lstlisting}

There are many ways the selection functions we have produced may be used by the community. The number density of \gaia sources in any coordinate system may be directly corrected to infer the true underlying distribution as demonstrated in \citet{Rix2021} for white dwarfs as a function of colour and absolute magnitude. One can take a forward modelling approach applying the selection function to a model and comparing with the published \gaia data then updating the model parameters to find the best fit. This will be the subject of two upcoming papers \citet{Everall2021a, Everall2021b}.

The selection functions can also be used to generate realistic mock \gaia catalogues from simulations similar to projects such as \citet{Rybizki2020} and Aurigaia \citep{Grand2018}. However, as discussed in Section~\ref{sec:discussion} features of the Milky Way are baked in due to crowding dependence of \gaia's source sensitivity so applying these selection functions to simulated galaxies will not perfectly reproduce the samples \gaia would record if it were observing the simulated galaxy.

\section{Conclusions}
\label{sec:conclusion}

We have evaluated the selection functions for three subsamples of the \gaia EDR3 data release: (i) Sources with published parallax and proper motion, (ii) the subset with $\mathrm{RUWE}<1.4$ and (iii) sources with DR2 radial velocities published with EDR3. 

 All selection functions are evaluated as a function of position on the sky using a spherical needlets model. The astrometry and RUWE selection functions are modelled as a function of $G$-band apparent magnitude and we construct selection functions for RUWE and RVS as a function of $G$ and $G-G_\mathrm{RP}$ colour with Gaussian Process priors.

  In all cases there is a strong dependence on crowding and features of the \gaia scanning law. For RVS, the effects of temperature cuts on the sample and selection on $G_\mathrm{RVS}$ using IGSL magnitudes heavily impacts the selection function.
 
 In all samples the model is able to reproduce the data down to scales $\sim 2$ degrees. The only areas in which the model breaks down are where the selection probability varies on smaller scales and our spatial model doesn't have the flexibility to reproduce the data. This is an issue in the Milky Way bulge and around globular clusters. Much of the spatial structure and magnitude and colour dependence of the selection functions is complicated, detailed and non-intuitive. This demonstrates the importance of empirically evaluating the selection functions for subsets of \gaia data.
 
 We provide a first-order estimate of the \gaia EDR3 source catalogue selection function to be used with the sub-sample selection functions we have developed. All of these are accessible through the \textsc{gaiaverse/selectionfunctions} GitHub repository. \footnote{\url{https://github.com/gaiaverse/selectionfunctions}}

\section*{Acknowledgements}
AE thanks the Science and Technology Facilities Council of
the United Kingdom for financial support. DB thanks Magdalen College for his fellowship and the Rudolf Peierls Centre for Theoretical Physics for providing office space and travel funds. This work has made use of data from the European Space Agency (ESA) mission \gaia (\url{https://www.cosmos.esa.int/gaia}), processed by the \gaia
Data Processing and Analysis Consortium (DPAC,
\url{https://www.cosmos.esa.int/web/gaia/dpac/consortium}). Funding for the DPAC has been provided by national institutions, in particular the institutions participating in the \gaia Multilateral Agreement.

AE is grateful to the \gaia helpdesk and in particular Lennart Lindegren and Paola Sartoretti for helping to explain key features of the selection functions. The authors also thank the anonymous referee for providing detailed comments which enhanced our descriptions of the selection functions.

\section*{Data availability}
The data underlying this article are publicly available from the European Space Agency's \gaia archive (\url{https://gea.esac.esa.int/archive/}). Open-source implementations of the selection functions calculated in this work are available through the Python package \textsc{selectionfunctions} (\url{https://github.com/gaiaverse/selectionfunctions}). The data products needed to evaluate the selection functions are publicly available on the Harvard Dataverse (\url{https://doi.org/10.7910/DVN/TCGI69}).




\bibliographystyle{mnras}
\bibliography{references} 




\appendix

\section{Binomial one-tailed p-value}
\label{app:binom}

\begin{figure}
  \centering
  \includegraphics[width=0.485\textwidth]{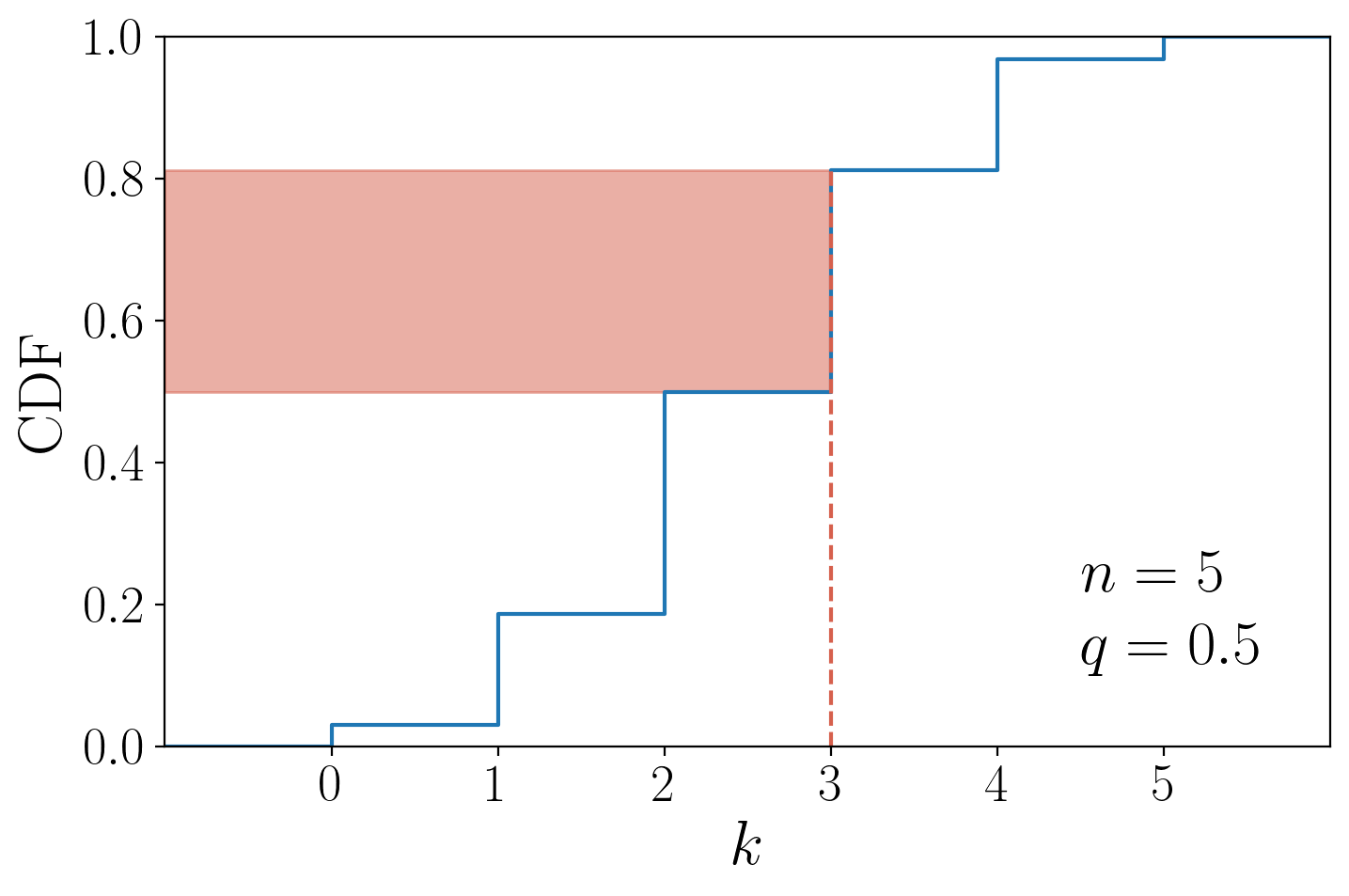}
  \caption[]{Changed p to q in this Fig The CDF of the Binomial distribution for $q=0.5$ and $n=5$ is shown with the blue line as a series of steps. The value of the CDF at $k=3$ is uniformly distributed between $0.50$ and $0.81$ shown by the red shaded region.}
   \label{fig:binom_example}
\end{figure}

The one-tailed p-value test provides the probability that the observations would be smaller (in some sense) than the actual measured data, $d$ given the hypothesised model.

For a hypothesised model with parameters $\psi$ and measured data $d$, the likelihood of the data given the parameters is $\prob(d\,|\,\psi)$. The p-value is given by the integral over this (i.e. the CDF)
\begin{equation}
    p_\mathrm{value} = \int_{d'_\mathrm{min}}^d \mathrm{d}d' \, \prob(d'\,|\,\psi)
\end{equation}
where $d'_\mathrm{min}$ is the minimum value the data can take under the model.

Let's say we have a sample of $k$ marbles drawn from a bag of $n$. We hypothesise that the probability of any marble being selected is Bernoulli (i.e. randomly) distributed with probability $q$. Therefore our likelihood is $\mathrm{Binomial}(k\,|\,n,q)$. The p-value is given by 
\begin{equation}
    p_\mathrm{value} = \int_{k'=0}^k \mathrm{d}k' \, \mathrm{Binomial}(k'\,|\,n,q).
\end{equation}
However, the CDF at $k'=k$ discontinuously jumps. This is shown by the example Binomial CDF in Fig.~\ref{fig:binom_example} where we have used $n=5$, $q=0.5$ to demonstrate. The CDF at $k=3$ has a discontinuity and uniformly covers a range of p-values in the CDF. Therefore we can write the p-value as 
\begin{equation}
    p_\mathrm{value} = \mathrm{U}\left[\sum_{k'=0}^{k-1} \mathrm{Binomial}(k'\,|\,n,q), \sum_{k'=0}^{k} \mathrm{Binomial}(k'\,|\,n,q)\right].
\end{equation}
This is shown by the red shaded region in Fig.~\ref{fig:binom_example} as the p-value for $k=3$ given $n=5$, $q=0.5$ is $p_\mathrm{value}\sim \mathrm{U}[0.50,0.81]$.

Without any data ($n=0$) the p-value becomes uniformly distributed
\begin{align}
    p_\mathrm{value} &= \mathrm{U}\left[\sum_{k'=0}^{-1} \mathrm{Binomial}(0\,|\,0,q), \sum_{k'=0}^{0} \mathrm{Binomial}(0\,|\,0,q)\right] \nonumber\\&= \mathrm{U}\left[0,1\right] .
\end{align}
As one would hopefully expect, if there's no data to test the hypothesis against, then the p-value is model independent (i.e. there's no dependence on $q$).

\section{\gaia EDR3 source catalogue}
\label{app:edr3}

\begin{figure}
  \centering
  \includegraphics[width=0.485\textwidth]{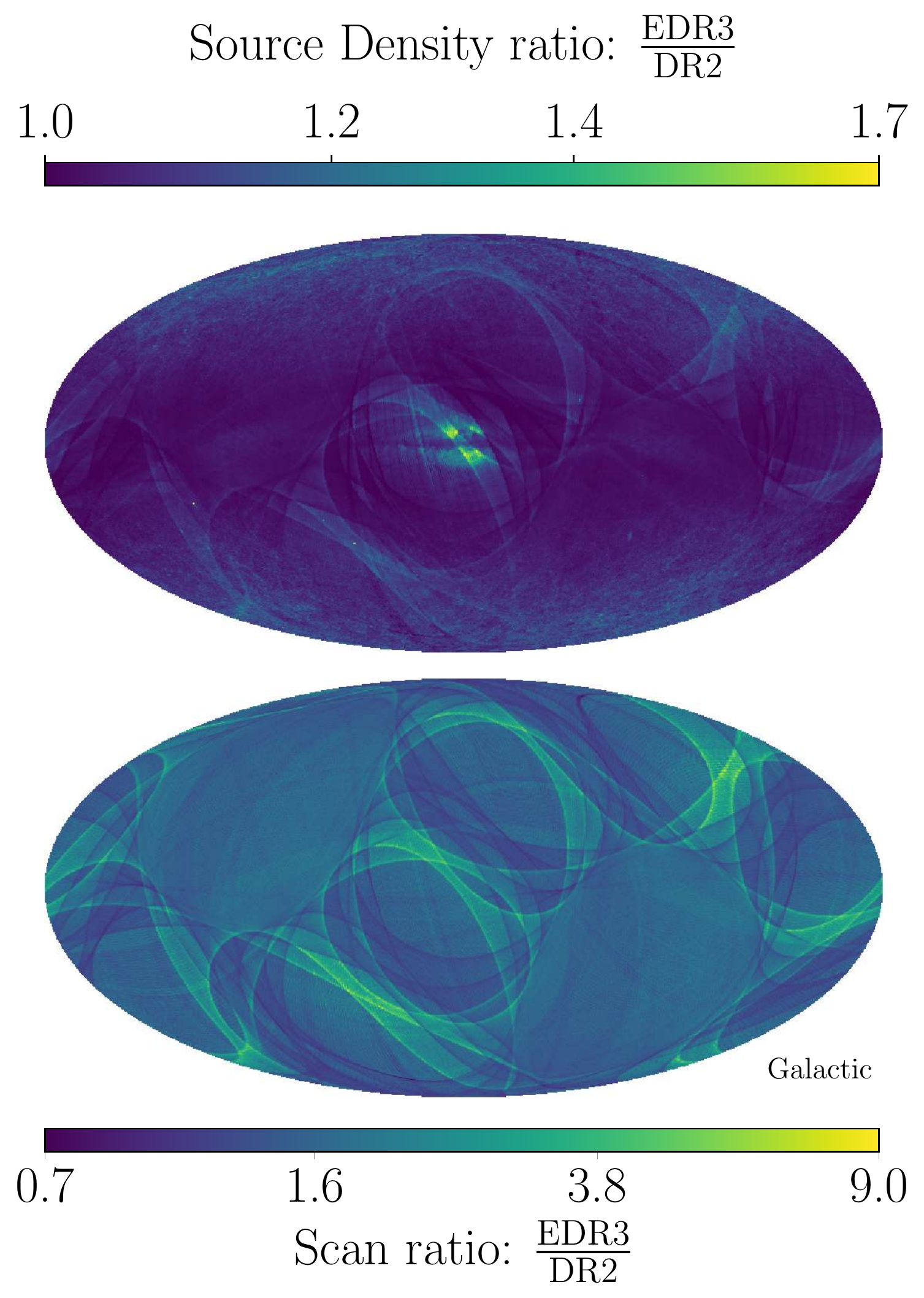}
  \caption[]{The increase of sources in the \gaia source catalogue between EDR3 and DR2 is driven by new scans. In the top panel we show the ratio of source density in HEALPix pixels between EDR3 and DR2 showing clear structure of the scanning law. The bottom panel is the approximate ratio of number of scans between the data releases using the calibrated DR2 scanning law and nominal scanning law for EDR3. Regions which received few observations in DR2 with many more in EDR3 saw large increase in number of objects in the \gaia source catalogue.}
   \label{fig:sd_ratio}
\end{figure}
\begin{figure*}
  \centering
  \includegraphics[width=\textwidth]{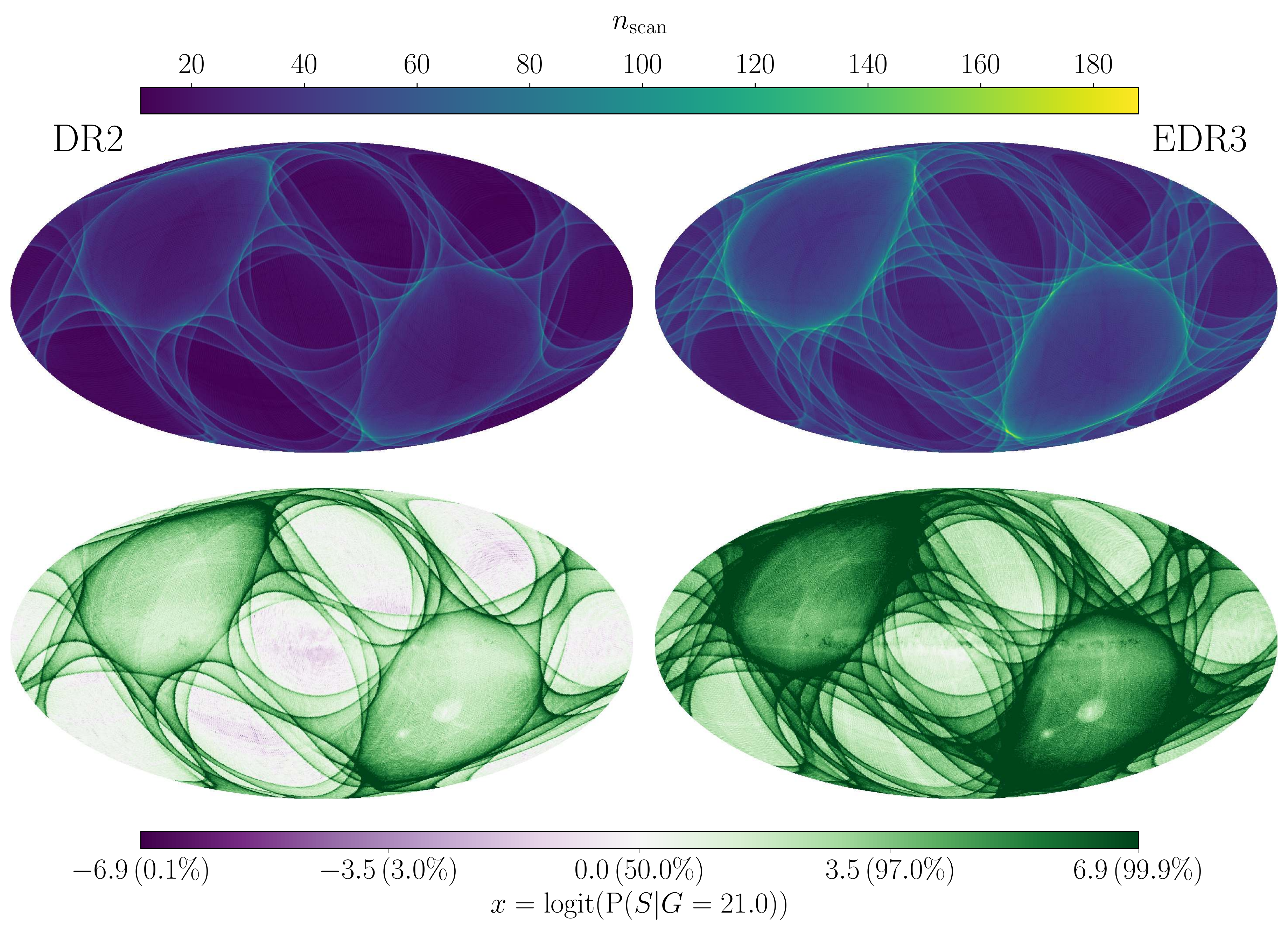}
  \caption[]{A simple estimate of the \gaia EDR3 selection function is achieved by updating the DR2 selection function with the nominal scanning law for EDR3. In the top row we show the number of scans received as a function of position on the sky in Galactic coordinates for DR2 (left) and EDR3 (right). The inferred \gaia source catalogue selection function probability at $G=21.0$ is shown in the bottom panel where regions which have received more scans in EDR3 have significantly higher selection probability.}
   \label{fig:edr3_sf}
\end{figure*}

This paper has focused on evaluating selection functions for subsamples of \gaia EDR3. However, these are only useful in conjunction with the \gaia source catalogue as described in Section~\ref{sec:method}.

In \citet{CoGII}, the \gaia DR2 source catalogue selection function was evaluated. The method involved evaluating the observation efficiency, the probability that a source transit would be used in the astrometric solution, for every source as a function of apparent magnitude and on-sky source density. The selection function was then the probability that a source received at least five successful transits. 

The method required precise predictions of the number of occasions on which the source could have been detected from the scanning law. The scanning law was calibrated using the \gaia DR2 epoch photometry \citep{CoGI, CoGIII}. However epoch photometry for the EDR3 baseline won't be published until the full data release in 2022 which prevents us from calibrating the EDR3 scanning law and repeating the method previously used for DR2.

Work is currently underway to produce a significantly more impressive selection function for the EDR3 source catalogue \citep{CoGVI, CoGVII}. For now, we provide a simple estimate for the EDR3 selection function based on \citet{CoGII} which can be used with the results of this paper but emphasise that this will be superseded when the new results are published.

The \gaia EDR3 source catalogue contains about $7$ per cent more sources than \gaia DR2. Fig~\ref{fig:sd_ratio} shows the ratio of source density across the sky between EDR3 and DR2. A striking feature of this is that the increase in source density follows patterns of the scanning law. Regions of the sky which were under-scanned in DR2 but received more transits in the extra 14 months of data to EDR3 have the biggest increase in content.

Motivated by this, we propose a very simple adjustment to evaluate the EDR3 selection function. We assume that the detection efficiency \gaia EDR3 is the same as in DR2. Therefore we use the results from Fig.~7 of \citet{CoGII} to evaluate the detection efficiency as a function of apparent magnitude and source density. The number of times a source was transited is evaluated from the uncalibrated \gaia EDR3 nominal scanning law. Therefore the selection function is \citep[Eq. C1][]{CoGII}
\begin{equation}
    \prob(\mathcal{S}_\mathrm{source}\,|\,G, l, b) = 1 - \sum_{m=0}^4 \binom{n}{m} \frac{\mathrm{Beta}(A + m, B + n - m)}{\mathrm{Beta}(A, B)}
\end{equation}
 where $A,B$ are the DR2 Beta distribution parameters of the efficiency as a function of source density and apparent magnitude. $n$ is the number of times a source was transited in EDR3 according to the EDR3 scanning law.
 
 The numbers of transits in DR2 and EDR3 from \citep{CoGIII} and the EDR3 nominal scanning law respectively are shown in the top panels of Fig.~\ref{fig:edr3_sf}. In the bottom panels we show the selection function probability at $G=21$. The selection probability has increased across the majority of the sky but most significantly in the ecliptic plane regions where there were few transits in DR2 such as the West side of the Galactic bulge.
 
 This result provides a first estimate of the \gaia EDR3 selection function but should be used with caution. There are three limitations to this model.
 
 Firstly, we have assumed that the pipeline of DR2 and EDR3 will lead to the same source detection efficiency in both catalogues. With each new data release, the data processing pipeline is rerun with improved source calibration which can change whether individual source observations are used in the astrometric solution.
 
 Secondly, in \gaia DR2, when two sources were observed to be within $0.4$ arcseconds of one another, the lower priority source was removed from the source catalogue. In EDR3, this threshold was reduced to $0.18$ arcseconds \citep{Brown2021, Torra2021} so EDR3 will be more complete in crowded regions independent of the scanning law.
 
 Finally, the EDR3 nominal scanning law has not been calibrated directly against the data and will deviate from \gaia's true scanning history by up to $\sim 30$ arcseconds at any point in time. Therefore the predicted observations in this model will be marginally off the true numbers over small regions of sky.
 
 Details on how to access the \gaia EDR3 source catalogue selection function are provided in Section~\ref{sec:access}.


\bsp	
\label{lastpage}
\end{document}